\documentclass[two column]{revtex4-1}
\date{today}
\usepackage{natbib,hyperref}
\usepackage{graphicx}
\usepackage{epstopdf}
\usepackage{mathrsfs}

\newcommand{\ket}[1]{\left|#1\right>}
\newcommand{\bra}[1]{\left< #1 \right|}

\newcommand{\beq}{\begin{equation}}
\newcommand{\eeq}{\end{equation}}
\newcommand{\beqa}{\begin{eqnarray}}
\newcommand{\eeqa}{\end{eqnarray}}

\begin{document}
\title{Spin relaxation of a donor electron coupled to interface states}
\author{Peihao Huang$^{1,2,3}$}
\email{phhuang.cmp@gmail.com}
\author{Garnett W. Bryant$^{1,2}$}
\affiliation{$^1$Joint Quantum Institute, National Institute of Standards and Technology and University of Maryland, Maryland, 20899}
\affiliation{$^2$Quantum Measurement Division, National Institute of Standards and Technology, Maryland, 20899}
\affiliation{$^3$Shenzhen Institute for Quantum Science and Engineering, and Department of Physics, Southern University of Science and Technology, Shenzhen 518055, China}

\date{\today}

\begin{abstract}
An electron spin qubit in a silicon donor atom is a promising candidate for quantum information processing because of its long coherence time. To be sensed  with a single-electron transistor, the donor atom is usually located near an interface, where the donor states can be coupled with interface states. Here we study the phonon-assisted spin-relaxation mechanisms when a donor is coupled to confined (quantum-dot-like) interface states. We find that both Zeeman interaction and spin-orbit interaction can hybridize spin and orbital states, each contributing to phonon-assisted spin relaxation in addition to the spin relaxation for a bulk donor or a quantum dot.
When the applied magnetic field $B$ is weak (compared to orbital spacing), the phonon assisted spin relaxation shows the $B^5$ dependence.
We find that there are peaks (hot-spots) in the $B$-dependent and detuning dependent spin relaxation due to strong hybridization of orbital states with opposite spin. We also find spin relaxation dips (cool-spots) due to the interference of different relaxation channels. Qubit operations near spin relaxation hot-spots can be useful for the fast spin initialization and near cool-spots for the preservation of quantum information during the transfer of spin qubits.

\end{abstract}

\maketitle

\section{introduction}

Electron spin qubits in semiconductor donors or quantum dots (QDs) are promising candidates for quantum information processing because of the tunability of electronic states and compatibility with existing semiconductor fabrication technologies.\cite{loss1998, kane1998, hanson2007, zwanenburg2013} Silicon as a host material for spin qubits is of particular interest because of the weak spin-orbit interaction (SOI) and the development of isotopic enrichment, which suppress both spin relaxation and pure dephasing. \cite{tyryshkin2003,morton2008,tyryshkin2012} Long coherence time and high fidelity readout have been demonstrated experimentally for a spin qubit in isotopically enriched silicon.\cite{muhonen2014} Furthermore, spin qubits in silicon donor atoms can be engineered by deterministic doping, where individual donor atoms can be placed with sub-nm precision by using scanning tunneling microscopy (STM) lithography. \cite{weber2014}

For readout with a single-electron transistor, the donor atoms of interest are usually located near an interface. \cite{morello2010,muhonen2014,singh2016} For example, in recent experiments with ion implanted phosphorus, the donors are separated about 10 nm from the interface. \cite{muhonen2014}
For these short separations, the interface states can couple to the donor state and alter the behavior of the spin qubit in a donor. The interface states are QD-like states confined laterally by the nearby donor potential and vertically by an applied electric field at the interface. The resulting lateral confinement along the interface can be as large as 10 meV. \cite{calderon2006, lansbergen2008}
It has been proposed to use the interface states to mediate dipole-dipole coupling between donor electrons\cite{tosi_silicon_2017}, or to transfer spin qubit information between remote donor atoms\cite{mohiyaddin2016}, where electrons are transferred between donors and interface states by applying an electric field. Recently, STM experiment demonstrated the tunnel coupling between donor and a QD-like state. \cite{salfi_valley_2018}


Spin relaxation describes how spin decays from its excited state to ground state, which is a type of decoherence that cannot be substantially suppressed with spin echo techniques. Spin relaxation  is an important quantity for the characterization of quantum systems, such as donor systems \cite{feher_electron_1959, tyryshkin2003, morton2008, morello2010, buch_spin_2013, dehollain_single-shot_2014, kolkowitz_probing_2015}, single QD system \cite{amasha_electrical_2008, hayes_lifetime_2009, yang2013, xiao_measurement_2010} and double QD (DQD) systems \cite{johnson_tripletsinglet_2005, pfund_suppression_2007, meunier_experimental_2007, prance_single-shot_2012}. In a single bulk donor, spin relaxation is dominated by the Zeeman interaction (ZI), which hybridizes the donor ground orbital and donor excited orbitals with opposite spin. \cite{roth1960,hasegawa1960,ramdas1981,tahan2002} In a single QD, SOI is the dominant mechanism for spin relaxation and hybridization of the QD ground state with the QD excited states of opposite spin. \cite{khaetskii2001, golovach2004, amasha_electrical_2008, hayes_lifetime_2009, yang2013,tahan2014,huang_spin_2014}
In tunnel coupled donor-interface system, hybridization of donor ground orbital with QD-like interface states could give rise to additional electron spin relaxation. A recent study of electron spin relaxation in a flip-flop qubit shows that the interface state can induce a strong spin relaxation peak (hot-spot) based on the single valley approximation. \cite{boross_valley-enhanced_2016} Spin relaxation hot-spots are also studied in GaAs double QD \cite{srinivasa_simultaneous_2013} and silicon QD \cite{yang2013}. However, there is still no comprehensive study of the spin relaxation in a donor coupled to a QD with the presence of the valley states in the QD. Furthermore, not much attention has been paid to the effect of destructive interference on spin relaxation.

In this work, we study the additional spin relaxation that arises when a donor is coupled to QD-like interface states. 
We find that both ZI and SOI can couple the donor ground state and QD states with opposite spin, and result in a phonon-assisted inter-donor-QD spin relaxation, in addition to the spin relaxation for a bulk donor or a quantum dot. A comprehensive comparison is done between the inter-donor-QD spin relaxation, the intra-donor spin relaxation, and the intra-QD spin relaxation. When the applied magnetic field $B$ is small, we find the phonon assisted spin relaxation always shows the $B^5$ dependence.
Multiple spin relaxation hot-spots are found due to the crossing of orbital states with opposite spin. We find spin relaxation cool-spots, where spin relaxation is suppressed due to the interference between different spin relaxation channels. The spin relaxation cool-spot happens when the spin-up state of the ground orbital (such as donor ground orbital) is between the spin-down states of the excited orbitals (such as the valley states of QD). The qubit operation near a spin relaxation hot-spot could be useful for the fast spin initialization and near a cool-spot for the preservation of quantum information during the coherent transfer of spin information between donor atoms via interface states.



The paper is organized as follows. In Sec. II, we set up the model Hamiltonian. In Sec. III, the effective spin-phonon interactions are obtained for two different spin hybridization mechanisms. In Sec. IV, the expressions of spin relaxation are given. In Sec. V, we show the numerical results of spin relaxation for various applied magnetic fields and detunings. In Sec. VI, we compare the result with spin relaxation in a single bulk donor and discuss possible consequences for experiments. Finally, we draw conclusions in Sec. VII.

\section{System Hamiltonian}

\begin{figure}
\centering
\includegraphics[scale=0.82, bb=0 0 238 193]{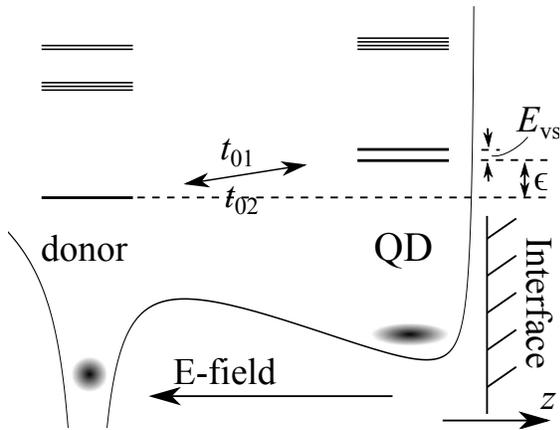}
\caption{Schematic diagram of system potential and energy levels of a donor coupled to a QD-like interface state under an electric field from metallic gates. The energy levels are shown for a donor and a QD confinement in the absence of tunnel coupling. $\epsilon$ is the detuning of QD and donor ground states, $E_{VS}$ is the valley splitting of the lowest two QD states,  $t_{01}$ and $t_{02}$ are the tunnel coupling amplitude between the donor ground state and the two lowest QD states.}\label{Fig_scheme}
\end{figure}

We consider the electron spin of a phosphorus (P) donor atom in an isotopically enriched $^{28}$Si with the P donor electron tunnel coupled to nearby QD-like interface states. Figure \ref{Fig_scheme} shows a schematic diagram of the electric potential and energy levels of the system. The system Hamiltonian is given by
\beq
H=H_O + H_Z + H_{SO} + H_{EP}, \label{H}
\eeq
where $H_O$ is the orbital part of the Hamiltonian (including the valley and envelope degrees of freedom), $H_Z$ is the ZI term in the presence of an applied magnetic field, $H_{SO}$ is the SOI in effective mass theory, and $H_{EP}$ is the electron-phonon interaction. We will describe each term in detail in the following paragraphs. 




The orbital Hamiltonian of a coupled donor-QD system can be expressed in the basis of donor eigenstates and QD eigenstates.
In silicon, the conduction band has six-fold degeneracy, known as valley degeneracy. The six valley states can be labeled as $x$, $\bar{x}$, $y$, $\bar{y}$, $z$, and $\bar{z}$.
In a bulk donor, the valley degeneracy split due to the donor confinement potential and the ground state has no valley degeneracy. In a QD, the potential at the interface splits the four $x$ and $y$ valleys from the two $z$ valley states, which are slightly split further. The energy splitting of the $z$ valley states is called valley splitting.
In this work, we are interested in the regime where the lowest two QD eigenstates are energetically close to the donor ground state. Since other donor or QD states are at least 10 meV higher, we will focus on the lowest three states: the ground state $\ket{0}$ of a single donor and the two lowest states $\ket{1}$ and $\ket{2}$ of the QD (see Appendix \ref{sec:EMA}).
%
Thus, the effective three-level Hamiltonian (suppressing the spin degree of freedom) is
\beqa
H_O&=&\epsilon_{0} \ket{0}\bra{0} + \epsilon_{1} \ket{1}\bra{1} + \epsilon_{2} \ket{2}\bra{2}  \nonumber\\
&&+ t_{01}\ket{0}\bra{1} + t_{02}\ket{0}\bra{2} + h.c.,
\eeqa
where $\epsilon_n$ (n=0, 1, 2) is the energy of each orbital basis state $\ket{n}$, $t_{01}$ and $t_{02}$ are the tunneling matrix elements between the donor ground state and the two QD states. The energies $\epsilon_n$ can be parameterized as $\epsilon_0=-\epsilon/2$, $\epsilon_1=\epsilon/2$ and $\epsilon_2=\epsilon/2+E_{VS}$, where $\epsilon$ is the detuning between donor and QD ground state, and $E_{VS}$ is the valley splitting between two QD states. The detuning $\epsilon$ is tunable with a metallic gate, and $E_{VS}$ ranges from tens of $\mu$eV to a few meV depending on interface potential and interface roughness. \cite{goswami_controllable_2007, shi_tunable_2011, yang2013, hao_electron_2014, veldhorst_spin-orbit_2015, gamble_valley_2016, kobayashi_resonant_2016}
In the presence of a magnetic field,
the ZI between the electron spin and the magnetic field is given by
\beq
H_Z=\frac{1}{2}\mu_B \sum_{j} P_j \vec{\sigma}\cdot \tensor{g}^{(j)} \cdot \vec{B},
\eeq
where $P_j$ is the projection operator that selects the $j$-th valley state (i.e. $j=x$, $\bar{x}$, $y$, $\bar{y}$, $z$, or $\bar{z}$), $\vec{\sigma}=(\sigma_x,\sigma_y,\sigma_z)$ is a vector of the Pauli matrices (z-axis is along [001] direction), and $\vec{B}=B(\sin\theta_B\cos\phi_B, \sin\theta_B\sin\phi_B, \cos\theta_B)$ is the applied magnetic field.
The anisotropic g-factor tensor $\tensor{g}^{(j)}$ is 
\beqa
\tensor{g}^{(j)} &=&g_{\perp} \tensor{1} + g_{ani} \tensor{U}^{(j)},
\eeqa
where $g_{\perp}$ ($g_{\parallel}$) is the g-factor perpendicular (parallel) to the valley ellipsoid, $g_{ani}=g_{\parallel}-g_{\perp}$ measures the extent of g-factor anisotropy, $\tensor{1}$ is an identity operator, and $\tensor{U}^{(j)}$ is an operator selecting the $|j|$-th direction. Note that, the g-factor tensor could in principle be different for donor and QD valley states. This more general form of ZI is considered in Appendix \ref{sec:Hz}.


The SOI couples the orbital and spin degrees of freedom and could affect the spin relaxation. The SOI due to the potential of the donor atom can be omitted since its effect on spin relaxation is small. \cite{hasegawa1960}
For a QD at an interface, we have the Rashba SOI,
\beq
H_{SO}=\alpha_{so} (p_{x}\sigma_y - p_{y}\sigma_x), \label{Hso}
\eeq
where $\alpha_{so}$ is the Rashba SOI constant, and $p_{x}$ and $p_{y}$ are the in-plane momentum operators (interface direction is assumed to be [001]).
Due to the bulk inversion symmetry of the silicon lattice, the Dresselhaus SOI in bulk silicon vanishes.
Note that, for a QD at an interface, Dresselhaus SOI may arise from the interface potential. \cite{nestoklon_electric_2008,prada_spinorbit_2011}
The presence of Dresselhaus SOI modifies the dependence of SOI induced spin relaxation on the orientation of the applied magnetic field, but not the dependence on the magnitude of the magnetic field. \cite{huang_electron_2014}
Furthermore, the ZI induced spin relaxation plays a more important role in most cases as shown below.
In this work, we consider only the Rashba SOI for simplicity.




The spin relaxation requires both a spin flip, which is mediated by the hybridization and energy dissipation.
The energy dissipation for relaxation is provided by the electron-phonon interaction $H_{EP}$. In silicon, we have (see Appendix \ref{sec:ep})
\begin{equation}
H_{EP}=\sum_j P_j \sum_{\vec{q}\lambda}e^{i\vec{q}\cdot\vec{r}} M_{\vec{q}\lambda}^{(j)} (b_{-\vec{q}\lambda}^\dag + b_{\vec{q}\lambda}), \label{Hep}
\end{equation}
\beq
M_{\vec{q}\lambda}^{(j)} =i\sqrt{\hbar q/2\rho_c v_{\lambda}} \Pi_{\vec{q}\lambda}^{(j)}, \label{Mqj}
\eeq
\beqa
{\Pi}_{\vec{q}\lambda}^{(j)}
&=& \hat{e}^{(\vec{q}\lambda)} \cdot (\Xi_{d}\tensor{1}+ \Xi_{u}\tensor{U}^{(j)}) \cdot \hat{q}, \label{Piqj}
\eeqa
where $P_j$ is the projection operator that selects the $j$-th valley,
$b_{\vec{q}\lambda}^\dag$ ($b_{\vec{q}\lambda}$) is the creation (annihilation) operator of a phonon with wave vector $\vec{q}$ and branch-index $\lambda$, $\lambda=l$ (longitudinal mode), $t_1$, or $t_2$ (transverse modes). $\rho_c$ is the sample density, $v_\lambda$ is phonon velocity, $\hat{e}^{(\vec{q}\lambda)}$ and $\hat{q}$ are unit vectors of phonon polarization and  wave vector, $ \Xi_{d}$ and $\Xi_{u}$ are the dilation and uniaxial shear deformation potential constants, and the coefficient $\Pi_{\vec{q}\lambda}^{(j)}$ is calculated and summarized in Appendix \ref{sec:ep}.


\section{Effective spin-phonon interaction}

The electron-phonon interaction can dissipate the energy of the electron, however, it does not relax spin without a mechanism that hybridizes the electron spin and orbital state.
To study spin relaxation, in principle, we should first solve $H_O + H_Z +H_{SO}$ to find the corresponding eigenstates, where states with different spins are hybridized. Then, by including electron-phonon interaction $H_{EP}$, one can calculate the relaxation of these eigenstates. However, because we are interested in the spin degree of freedom, and because the hybridization of spin and orbit is small, one can treat this problem by perturbation theory without losing accuracy. Suppose the orbital Hamiltonian is diagonalized as $H_O=E_{\bar{n}}\ket{\bar{n}}\bra{\bar{n}}$,
\beq
\ket{\bar{n}}=\sum_{n} C_{\bar{n}n} \ket{n}, \label{psi}
\eeq
where $E_{\bar{n}}$ and $\ket{\bar{n}}$ are the orbital eigenenergies and eigenstates, and the orbital basis state $\ket{n}$ can be expressed in terms of envelope and Bloch functions (see Appendix \ref{sec:EMA}). Then, we consider $H=H_0+H_{h} + H_{EP}$, where $H_0=H_O + \sum_{\bar{n}} \bra{\bar{n}}H_Z\ket{\bar{n}}\ket{\bar{n}}\bra{\bar{n}}$ is the unperturbed part, and $H_{h} + H_{EP}$ is the perturbation. $H_{h}=\sum_{\bar{n}\ne \bar{n}^\prime}\left[\bra{\bar{n}}H_Z\ket{\bar{n}^\prime} \ket{\bar{n}}\bra{\bar{n}^\prime} + h.c.\right] + H_{SOI}$ hybridizes spin and orbital state, and $H_{EP}$ provides energy dissipation of the system.
To 2nd order of $H_{h} + H_{EP}$, the effective spin-flip Hamiltonian is
\beq
(H)_{\bar{0}\bar{0}}^{\uparrow\downarrow} =\frac{1}{2} \sum_{\bar{n}\ne \bar{0}} \bigg\{ \frac{(H_{h})_{\bar{0}\bar{n}}^{\uparrow \downarrow} (H_{EP})_{\bar{n}\bar{0}}^{\downarrow\downarrow}}{E_{\bar{0}}-E_{\bar{n}} + E_Z}
  + \frac{(H_{EP})_{\bar{0}\bar{n}}^{\uparrow\uparrow} (H_{h})_{\bar{n}\bar{0}}^{\uparrow\downarrow }}{E_{\bar{0}}-E_{\bar{n}}-E_Z} \bigg\},
 \label{H100} 
\eeq
where $(H)_{\bar{n}\bar{n}^\prime}^{ss^\prime}\equiv\bra{\bar{n}s} H \ket{\bar{n}^\prime s^\prime}$ and $E_Z$ is the ground orbital Zeeman splitting determined by $\bra{\bar{0}}H_Z\ket{\bar{0}}$.

In the following subsections, we will obtain the matrix elements of $H_{Z}$, $H_{SO}$ and $H_{EP}$. We will use a simplified notation $(H_i)_{\bar{n}\bar{n}^\prime}\equiv\bra{\bar{n}} H_i \ket{\bar{n}^\prime }$ for the matrix element in the diagonalized orbital basis, and $H_{i,{n}{n}^\prime}\equiv\bra{{n}} H_i \ket{{n}^\prime }$ for the matrix element in the original orbital basis, where $H_i$ can be $H_{Z}$, $H_{SO}$ or $H_{EP}$. We also use $\bar{r}$ to denote the excited orbital states, i.e. $\bar{r}\ne \bar{0}$ is always used.

\subsection{ZI Induced Hybridization}

ZI induced spin relaxation is known to be the dominant spin relaxation mechanism in a single donor, where g-factor anisotropy leads to  hybridization of spin-valley states.\cite{roth1960,hasegawa1960}
In a coupled donor-QD system, ZI will also hybridize spin and orbital states $\ket{\bar{0}}$, $\ket{\bar{1}}$ and $\ket{\bar{2}}$, and give rise to additional spin relaxation besides the relaxation in a single donor.

To find the matrix element of ZI in the orbital eigenstates, we first express ZI in the basis of donor ground state $\ket{0}$ and QD ground states $\ket{1}$ and $\ket{2}$. Then, ZI is
\beqa
H_{Z,nn^\prime} &=& \frac{1}{2}\mu_B \vec{\sigma}\cdot  \tensor{g}^{(nn^\prime)} \cdot \vec{B}, \\
\tensor{g}^{(nn^\prime)} &=& \left(g_{\perp} \tensor{1} + g_{ani} \tensor{D}^{(nn)}\right) \delta_{nn^\prime}, \\
\tensor{D}^{(nn^\prime)} &=& \sum_{j} \alpha_{n}^{(j)}\alpha_{n^\prime}^{(j)} \tensor{U}^{(j)},
\eeqa
where $\alpha_{n}^{(j)}$ is probability amplitude of state $\ket{n}$ in the $j$-th valley.
Then, we can express ZI in the basis of orbital eigenstates $\ket{\bar{n}}$. From Eq. (\ref{psi}), $\ket{\bar{n}}=\sum_{n} C_{\bar{n}n}\ket{{n}}$,
ZI in the basis of orbital eigenstates is 
\beqa
(H_Z)_{\bar{0}\bar{n}} &=& \frac{1}{2}\mu_B \vec{\sigma}\cdot  \tensor{G}^{(\bar{0}\bar{n})} \cdot \vec{B},
\eeqa
\beqa
\tensor{G}^{(\bar{0}\bar{n})} = g_{\perp}\tensor{1} \delta_{\bar{0}\bar{n}} + g_{ani}\sum_{n}  C_{\bar{0}n}^*C_{\bar{n}n} \tensor{D}^{(nn)},
\eeqa
where the off-diagonal elements of $\tensor{G}^{(\bar{0}\bar{n})}_{ij}$ are zero, and the diagonal elements are
\beqa
\tensor{G}^{(\bar{0}\bar{n})}_{xx}=\tensor{G}^{(\bar{0}\bar{n})}_{yy} &=&g_{\perp} \delta_{\bar{0}\bar{n}} + g_{ani}C_{\bar{0}0}^*C_{\bar{n}0}\tensor{\Delta}_{xx},\\
\tensor{G}^{(\bar{0}\bar{n})}_{zz}&=&g_{\parallel} \delta_{\bar{0}\bar{n}} + g_{ani} C_{\bar{0}0}^*C_{\bar{n}0} \tensor{\Delta}_{zz},
\eeqa
where the orthogonal relations $\sum_n C_{\bar{0}n}^*C_{\bar{n}n}=\delta_{\bar{0}\bar{n}}$ have been employed and then tensor $\tensor{\Delta}$ is
\beq
\tensor{\Delta} = \tensor{D}^{(00)} - \tensor{D}^{(11)} = \frac{1}{3}\tensor{1} - \tensor{U}^{(z)}.
\eeq

To find the effective spin-orbit hybridization term, we need to express ZI in a new (X,Y,Z) coordinate system, where $Z$ axis is along the spin quantization axis determined by $(H_Z)_{\bar{0}\bar{0}}$, so that $(H_Z)_{\bar{0}\bar{0}}=\frac{1}{2}E_Z \sigma_Z$, where $E_Z = g_{eff} \mu_B B$ is the Zeeman splitting and $g_{eff} =\sqrt{(\tensor{G}^{\bar{0}\bar{0}}_{xx}\sin\theta_B)^2+ (\tensor{G}^{\bar{0}\bar{0}}_{zz}\cos\theta_B)^2}$ is the effective g-factor [see Appendix \ref{sec:Hz}]. The spin quantization axis is different from the direction of $B$ because of the anisotropy of $G_{\xi\xi}^{\bar{0}\bar{0}}$. However, since
$g_{\perp}\approx g_{\parallel}\approx 2 \gg g_{ani}$,
the spin quantization can be taken approximately along the applied magnetic field.
Therefore, in the new $(X,Y,Z)$ coordinate, the spin-orbit hybridization term due to ZI is
\beqa
(H_Z)_{\bar{0}\bar{r}}^{\uparrow\downarrow} &\equiv&  \frac{1}{2} \mu_B B  g_{X}^{\bar{0}\bar{r}} \sigma_{X}^{\uparrow\downarrow},
\eeqa
\beqa
g_X^{\bar{0}\bar{r}} 
&\approx&g_{ani}C_{\bar{0}0}^*C_{\bar{r}0}\tensor{\Delta}_{XZ},
\label{Hzbarnbarn}
\eeqa
where $\sigma_{X}^{\uparrow\downarrow}=1$ and $\tensor{\Delta}_{XZ}=  \frac{1}{2} \sin2\theta_B$.

Eq. (\ref{Hzbarnbarn}) indicates that the hybridization due to ZI is proportional to $C_{\bar{0}0}^*C_{\bar{r}0}$, the g-factor anisotropy $g_{ani}$, and $\tensor{\Delta}_{XZ}=  \frac{1}{2} \sin2\theta_B$. The hybridization is maximum when $\theta_B=45^\circ$, and is zero when $\theta_B=0^\circ$ or $90^\circ$. The hybridization becomes zero because the spin quantization direction (given by $\tensor{G}^{\bar{0}\bar{0}}\cdot\vec{B}$) and the direction of spin operator (given by $\tensor{G}^{\bar{0}\bar{r}}\cdot\vec{B}$) in the hybridization term $(H_Z)_{\bar{0}\bar{r}}$ are along the same direction as $\vec{B}$, when $\vec{B}$ is along the main axis of g-factor tensor.
In this case, the transverse coupling of spin to phonon, which is responsible for spin relaxation, becomes zero.

Finally, if g-factor tensor $\tensor{g}^{(j)}$ is considered to be different between donor and QD, one can show that the only difference in the hybridization term is that $g_{ani}$ is replaced by $g_{QD,ani}$, which is the g-factor anisotropy in the QD (see Appendix \ref{sec:Hz}).

\subsection{SOI Induced Hybridization}

The SOI can also hybridize the spin and orbital states. Together with the electron-phonon interaction, it will induce spin relaxation.
To evaluate the matrix element $(H_{SO})_{\bar{0}\bar{r}}^{\uparrow\downarrow}$, i.e.
$(p_x)_{\bar{0}\bar{r}}$ and $(p_y)_{\bar{0}\bar{r}}$, it is convenient to use the commutation relation $[x,H_O]\approx i\hbar p_x/m^*$, where an average effective mass $m^*= 0.315 m_0$ is chosen because of the presence of multiple valley states (see Appendix \ref{sec:commutation}). From the commutation relation, we have
\beq
(p_x)_{\bar{0}\bar{r}} \approx m^*E_{\bar{r}\bar{0}}x_{\bar{0}\bar{r}}/(i\hbar),
\eeq
where $x_{\bar{0}\bar{r}} = \bra{\bar{0}}x\ket{\bar{r}}$ and $E_{\bar{r}\bar{0}}=E_{\bar{r}}-E_{\bar{0}}$ is the energy difference of the eigenstates in the absence of SOI. By using the single effective mass $m^*$, the estimated matrix element could be different from actual values by at most a factor of three.

The matrix element $x_{\bar{0}\bar{r}}$ can be written as
\beqa
x_{\bar{0}\bar{r}}&=& \sum_{nn^\prime} C_{\bar{0}n}^*C_{\bar{r}n^\prime} x^{nn^\prime}
=x_{\bar{0}\bar{r}}^{(1)}+x_{\bar{0}\bar{r}}^{(2)}, \label{x0r}
\eeqa
where $x^{nn^\prime} =\bra{n} x\ket{n^\prime}$ is the matrix element in the original basis, and $x_{\bar{0}\bar{r}}^{(1)} $ ($x_{\bar{0}\bar{r}}^{(2)} $) contains only the terms with $n^\prime=n$ ($n^\prime\neq n$) in the summation.
In $x_{\bar{0}\bar{r}}^{(2)}$, the terms of $x^{01}$ and $x^{02}$ are small because of the spatial separation of donor and QD states. The only term could contribute to $x_{\bar{0}\bar{r}}^{(2)}$ is $x^{12}$ that couples two valley states of QD, which can result in an inter-valley spin relaxation. The magnitude of $x^{12}$ is on the order of 1 nm, as estimated in a recent experiment,\cite{yang2013} and it should be even smaller for a flat interface. The term $x_{\bar{0}\bar{r}}^{(2)}$ will be omitted because its magnitude is small compared to $x_{\bar{0}\bar{r}}^{(1)}$ as shown below. (Note that this inter-valley spin relaxation is different from the intra-QD inter-valley spin relaxation, which is discussed later in Sec. V.B)

For the term $x_{\bar{0}\bar{r}}^{(1)}$, one can show that $x_{\bar{0}\bar{r}}^{(1)}=C_{\bar{0}0}^*C_{\bar{r}0}d_x$, where $d_x$ is the projection of a vector connecting the center of donor and center of QD on the x-axis. Thus,
\beqa
(p_x)_{\bar{0}\bar{r}}&\approx&
 C_{\bar{0}0}^*C_{\bar{r}0} (a_{so}/ \alpha_{so}) E_{\bar{r}\bar{0}}\cos\phi_d,
\eeqa
where $a_{so}= \alpha_{so} m^* d_{\parallel}/(i\hbar) $, $d_{\parallel}$ is the in-plane separation between donor and QD, $\phi_d$ is introduced so that $d_x=d_{\parallel}\cos\phi_d$ and $d_y=d_{\parallel}\sin\phi_d$. Therefore, from Eq. (\ref{Hso}),
\beqa
(H_{SO})_{\bar{0}\bar{r}}^{\uparrow\downarrow}&=& C_{\bar{0}0}^*C_{\bar{r}0} a_{so}E_{\bar{r}\bar{0}}\sigma_{x^{\prime\prime}}^{\uparrow\downarrow},
\eeqa
where $\sigma_{x^{\prime\prime}}^{\uparrow\downarrow}=(\sigma_y^{\uparrow\downarrow} \cos \phi_d - \sigma_x^{\uparrow\downarrow} \sin\phi_d)$.
Similarly, we have
\beqa
(H_{SO})_{\bar{r}\bar{0}}^{\uparrow\downarrow}
&=&-C_{\bar{r}0}^*C_{\bar{0}0} a_{so}E_{\bar{r}\bar{0}}\sigma_{x^{\prime\prime}}^{\uparrow\downarrow},
\eeqa
where the minus sign indicates that there will be cancellation of two terms in Eq. (\ref{H100}) in the limit of zero magnetic field. This cancellation, known as Van-Vleck cancellation\cite{van_vleck_paramagnetic_1940, orbach_theory_1961}, will result in an extra $E_Z^2$ dependence (besides the contribution of $E_Z^3$ from phonon spectral density) for spin relaxation. The results also indicate that the hybridization due to SOI is propotional to $a_{so}=\alpha_{so} m^* d_{\parallel}/(i\hbar) $. Thus, it is proportional to SOI strength $\alpha_{so}$ and the lateral separation of the donor and QD. Interestingly, there is no coupling if the donor and QD are vertically aligned. In that case, $(p_x)_{\bar{0}\bar{r}}$ and $(p_x)_{\bar{0}\bar{r}}$ are zero due to the vanishing of $x_{\bar{0}\bar{r}}^{(1)}$.
In this work, we assume a finite $d_{\parallel}$, so that the dominant contribution is from $x_{\bar{0}\bar{r}}^{(1)}$ rather than $x_{\bar{0}\bar{r}}^{(2)}$. 

\subsection{Electron-Phonon Matrix Elements}

The energy dissipation is provided by the electron-phonon interaction, which couples the ground $\ket{\bar{0}}$ and excited orbital states $\ket{\bar{r}}$ (i.e. $\bar{r}\ne \bar{0}$) with the same spin orientation. In this subsection, we calculate the matrix elements $\bra{\bar{0}} H_{EP} \ket{\bar{r}}$ of the electron-phonon interaction.
From Eq. (\ref{Hep}), (\ref{Mqj}), (\ref{Piqj}) and (\ref{psi}), we have
\beq
(H_{EP})_{\bar{0}\bar{r}}^{\uparrow\uparrow} = \sum_{\vec{q}\lambda} (M_{\vec{q}\lambda})_{\bar{0}\bar{r}} (b_{-\vec{q}\lambda}^\dag + b_{\vec{q}\lambda}),
\eeq
\beq
(M_{\vec{q}\lambda})_{\bar{0}\bar{r}} = \sum_{nn^\prime} C_{\bar{0}n}^*C_{\bar{r}n^\prime}  M_{\vec{q}\lambda,nn^\prime} ,
\eeq
\beq
M_{\vec{q}\lambda,nn^\prime}=\sum_j \alpha_{n}^{(j)}\alpha_{n^\prime}^{(j)}f_{nn^\prime }^{(j)}(\vec{q}) M_{\vec{q}\lambda}^{(j)},
\eeq
where $(H_{EP})_{\bar{0}\bar{r}} \equiv \bra{\bar{0}}  H_{EP} \ket{\bar{r}}$,
and the form factor $f_{nn^\prime}^{(j)}(\vec{q})$ is
\beq
f_{nn^\prime}^{(j)}(\vec{q}) \equiv \int d\vec{r} F_{nj}^*(\vec{r}) F_{n^\prime j}(\vec{r}) e^{i\vec{q}\cdot\vec{r}}, 
\eeq
where $F_{nj}^*(\vec{r}) $ is the envelope function in the $j$-th valley.
For spin splitting around $\omega_Z\equiv E_Z/\hbar=1$ GHz, we have $|q|=\omega_Z/v_j\approx 1$ $\mu m ^{-1}$. Thus, the wavelength of phonons which are on resonance with the spin splitting is much larger than the geometrical size of the donor-QD system ($\sim$ 10 nm).
In the limit of long wave phonons, we have $\exp({i\vec{q}\cdot\vec{r}}) \approx 1$ and $f_{nn^\prime}^{(j)}(\vec{q})\approx \delta_{nn^\prime}$.
Therefore,
\beq
M_{\vec{q}\lambda,nn^\prime}=\sum_j |\alpha_{n}^{(j)}|^2M_{\vec{q}\lambda}^{(j)},
\eeq
%
where $M_{\vec{q}\lambda,22}=M_{\vec{q}\lambda,11} \ne M_{\vec{q}\lambda,00}$. Because of the orthogonal relation $\sum_n C_{\bar{0}n}^*C_{\bar{r}n}=0$, we have
\beq
(M_{\vec{q}\lambda})_{\bar{0}\bar{r}}
= C_{\bar{0}0}^*C_{\bar{r}0} M_{\vec{q}\lambda,00}^{\prime},
\eeq
\beqa
M_{\vec{q}\lambda}^\prime &=&
i\sqrt{\hbar q/2\rho_c v_{\lambda}}\Pi_{\vec{q}\lambda}^{\prime},
\eeqa
\beqa
{\Pi}_{\vec{q}\lambda}^\prime
&=&\Xi_{u}\hat{e}^{(\vec{q}\lambda)} \cdot \tensor{\Delta} \cdot \hat{q},
\eeqa
where $M_{\vec{q}\lambda}^\prime = M_{\vec{q}\lambda,00} - M_{\vec{q}\lambda,11}$, $\Pi_{\vec{q}l}^{\prime} = \Xi_u (1/3-\cos^2\vartheta)$, $\Pi_{\vec{q}t_1}^{\prime}= \Xi_u \cos\vartheta\sin\vartheta$, and $\Pi_{\vec{q}t_2}^{\prime}=0$ (see Appendix \ref{sec:ep}). Therefore, in the long wave limit, the electron-phonon interaction matrix elements between orbital eigenstates is proportional to the extent of mixing of donor and QD states $C_{\bar{0}0}^*C_{\bar{r}0}$, and proportional to the uniaxial shear deformation potential constant $\Xi_u$.

The matrix element $M_{\vec{q}\lambda}^\prime \propto q^{1/2}$ is similar to the case of an electron in a bulk donor, but  different from the case of a single QD. For an electron in a single QD, the electron-phonon interaction matrix element scales as $q^{3/2}$, where the extra $q$ is because of the dipole interaction needed to couple the ground and excited states. Thus, for the same hybridization mechanism of SOI, the phonon-induced spin relaxation in a QD has an extra $B^2$ dependence compare to the spin relaxation studied here. \cite{khaetskii2001, golovach2004}

\subsection{Summary of Effective Spin-Phonon Interactions} 

With the form of electron-phonon interaction matrix elements, the effective spin-phonon Hamiltonian is
\beqa
(H^\prime)_{\bar{0}\bar{0}}^{\uparrow\downarrow}
&=& \frac{H_{EP}^\prime}{2}\sum_{r} \frac{- E_{\bar{r}\bar{0}} H_{h}^+ + E_Z H_{h}^-}{E_{\bar{r}\bar{0}}^2-E_Z^2} ,
\eeqa
\beqa
H_{EP}^{\prime} &=& \sum_{\vec{q}\lambda} M_{\vec{q}\lambda}^\prime (b_{-\vec{q}\lambda}^\dag + b_{\vec{q}\lambda}),
\eeqa
\beqa
H_{h}^+&=&C_{\bar{r}0}^*C_{\bar{0}0} (H_{h})_{\bar{0}\bar{r}}^{\uparrow \downarrow}  + C_{\bar{0}0}^*C_{\bar{r}0} (H_{h})_{\bar{r}\bar{0}}^{\uparrow\downarrow },\\
H_{h}^-&=&C_{\bar{r}0}^*C_{\bar{0}0} (H_{h})_{\bar{0}\bar{r}}^{\uparrow \downarrow}  - C_{\bar{0}0}^*C_{\bar{r}0} (H_{h})_{\bar{r}\bar{0}}^{\uparrow\downarrow }.
\eeqa
There are two possible hybridization mechanisms, i.e. $H_Z$ and $H_{SO}$. The interference between different mechanisms can be neglected since the spin relaxation of these two mechanisms are in general of different magnitude; Furthermore, there can always be a phase difference between the effective spin-phonon interaction of different mechanisms. In the following, we will study separately each hybridization mechanism and neglect the interference of the two hybridization mechanisms.

When $H_{h}=H_{Z}$, $H_{h}^- $ vanishes due to the same sign and magnitude of matrix elements of $(H_{Z})_{\bar{0}\bar{r}}^{\uparrow\downarrow}$ and $(H_{Z})_{\bar{r}\bar{0}}^{\uparrow\downarrow}$; While
\beqa
H_{h}^+&=& |C_{\bar{r}0}^*C_{\bar{0}0}|^2 g_{ani} \tensor{\Delta}_{XZ} \mu_BB,
\eeqa
where $\sigma_{X}^{\uparrow\downarrow}=1$ in the (X,Y,Z) coordinate.
Therefore, the effective spin-flip Hamiltonian due to ZI and electron-phonon interaction is
\beq
(H_Z+H_{EP})_{\bar{0}\bar{0}}^{\uparrow\downarrow}=-\frac{g^\prime }{2} \tensor{\Delta}_{XZ}  \eta_{Z} E_Z H_{EP}^\prime, \label{Heff_ZI} 
\eeq
\beq
\eta_{Z}\equiv \sum_{r=1,2} \frac{E_{\bar{r}\bar{0}}|C_{\bar{0}0}^*C_{\bar{r}0}|^2}{E_{\bar{r}\bar{0}}^2 - E_Z^2}, \label{eta_Z}
\eeq
where $g^\prime\equiv g_{ani}/g_{eff}$ is the rescaled g-factor anisotropy and the coefficient $\eta_{Z}$ accounts for contributions from different orbitals to the effective spin-phonon interaction.


When $H_{h}=H_{SO}$, $H_{h}^+ $ vanishes due to the different signs of matrix elements of $(H_{SO})_{\bar{0}\bar{r}}^{\uparrow\downarrow}$ and $(H_{SO})_{\bar{r}\bar{0}}^{\uparrow\downarrow}$; While
\beqa
H_{h}^-
&=& 2|C_{\bar{r}0}^*C_{\bar{0}0}|^2 a_{so}\sigma_{x^{\prime\prime}}^{\uparrow\downarrow} E_{\bar{r}\bar{0}}.
\eeqa
Therefore, the effective spin flip Hamiltonian due to SOI and electron-phonon interaction is 
\beq
(H_{SO} + H_{EP})_{\bar{0}\bar{0}}^{\uparrow\downarrow}=a_{so}\sigma_{x^{\prime\prime}}^{\uparrow\downarrow}\eta_{SO} E_Z H_{EP}^\prime, \label{Heff_SO}
\eeq
where $\eta_{SO}=\eta_{Z}$.

$H_Z$ and $H_{SO}$ are different and the cancellation of terms is different, however, the final effective spin-phonon interaction Hamiltonians are similar. The g-factor anisotropy in ZI is a result in part of the microscopic SOI, so both mechanisms originate from SOI, one from microscopic SOI not in the effective mass theory, the other from the SOI in the effective mass theory. Both hybridizations show the same dependence with $E_Z$. They also show the same dependence on the hybridization of orbital states $\eta_{SO}=\eta_Z$. The only difference is the angular dependencies with magnetic field due to the difference between $\tensor{\Delta}_{XZ}$ and $\sigma_{x^{\prime\prime}}^{\uparrow\downarrow}$.

The strength of spin-phonon interaction for both mechanisms is proportional to $\eta_Z$, which depends on the energies of spin-orbital states.
$\eta_Z$ can be strongly enhanced, when the spin state is in resonant with the orbital states, where a corresponding spin relaxation hot-spot appears. $\eta_Z$ can also be zero, when contributions from different orbitals cancel with each other, where spin relaxation is strongly suppressed (spin relaxation cool-spot).

\section{Spin relaxation}

During the setup of the Hamiltonian, we consider only one state in the donor, and two valley states in the QD. In this case, we implicitly omit the intra-donor spin relaxation and intra-QD spin relaxation, in which the excited donor states and the excited QD states are involved. Thus, the spin relaxation based on Hamiltonian (\ref{H}) will be called inter-donor-QD spin relaxation. In this section, we first derive the expressions of inter-donor-QD spin relaxation. Then, we discuss the intra-donor and intra-QD spin relaxation mechanisms. The total spin relaxation will be the summation of contributions of  all the spin relaxation mechanisms. 


\subsection{Inter-donor-QD spin relaxation}

In this subsection, we study the inter-donor-QD spin relaxation due to the effective spin-phonon interaction Eq. (\ref{Heff_ZI}) and Eq. (\ref{Heff_SO}).
The spin relaxation time is given by $1/T_1=W_{\uparrow\downarrow} + W_{\downarrow\uparrow}$, where $W_{\downarrow\uparrow}=\Gamma [n(\omega_{\vec{q}\lambda})+1]$ is the rate for transition from the higher-energy (spin-up) state to lower energy (spin-down) state (emitting phonon) and $W_{\uparrow\downarrow}=\Gamma_1 n(\omega_{\vec{q}\lambda})$ is the rate for the opposite transition (absorbing phonon),
where
$
\Gamma_1= \frac{2\pi}{\hbar} \sum_{\vec{q}\lambda} |(H)_{\bar{0}\bar{0}}^{\uparrow\downarrow}|^2  \delta (E_Z - \hbar \omega_{\vec{q}\lambda}).
$
Considering the zero temperature limit, the inter-donor-QD spin relaxation is given by
\beq
1/T_{1,inter-donor-QD} = \Gamma_{Z-ph} + \Gamma_{SO-ph}
\eeq
where $\Gamma_{Z-ph}$ is the spin relaxation due to ZI and phonon noise, and $\Gamma_{SO-ph}$ is the spin relaxation due to SOI and phonon noise. We find that
\beqa
\Gamma_{Z-ph} &=& \frac{1}{4}\left({g^{\prime}\eta_{Z}\tensor{\Delta}_{XZ} }\right)^2 (\hbar\omega_Z )^2\Gamma_{ph}(\omega_Z),\\
\Gamma_{SO-ph} &=&  \left|a_{so}\eta_{SO} \sigma_{x^{\prime\prime}}^{\uparrow\downarrow} \right|^2 (\hbar\omega_Z)^2 \Gamma_{ph}(\omega_Z),
\eeqa
\beqa
\Gamma_{ph}(\omega_Z) &=& \frac{1}{4\pi\rho_c\hbar}\sum_{\lambda} \frac{\omega_Z^3}{v_\lambda^5} \int_{0}^{\pi} d\vartheta \sin\vartheta \Pi_{\vec{q}\lambda}^{\prime 2}(\vartheta),
\eeqa
where $\omega_Z=E_Z/\hbar$ is the electron Zeeman frequency.

The phonon temperature is essentially zero since the temperature in a dilution refrigerator is around 10 mK, so that the phonon thermal energy is much less than the electron Zeeman splitting. With an elevated phonon temperature, the number of phonon excitation increases, then, an extra factor $\coth (\hbar \omega_Z/k_B T)$ appears in the expression of spin relaxation, which can play a role when thermal energy $k_BT$ is comparable to the Zeeman energy $\hbar\omega_Z$. Also, with an elevated phonon temperature, the phonon coherence length and coherence time can be shorter, and the destructive interference effect discuss below can be broadened.

The analytical expressions of the inter-donor-QD spin relaxation indicate that both spin relaxation mechanisms show the same $B^5$ (or $\omega_Z^5$) dependence on the magnitude of the applied magnetic field. They also show the same dependence on the hybridization of orbital states since $\eta_{SO}=\eta_Z$. The two mechanisms show different angular dependencies on the orientation of the applied magnetic field. They also show different magnitudes depending on other parameters. Spin relaxation due to ZI shows $g_{ani}^2$ dependence on the g-factor anisotropy, while spin relaxation due to ZI shows $\alpha_{so}^2$ dependence on the SOI strength and $|d_{\parallel}|^2$ dependence on the lateral separation of the donor and the QD. The spin relaxation due to ZI shows $\tensor{\Delta}_{XZ}^2=\sin^2(2\theta_B)$ dependence on the orientation of the applied magnetic field, while, the spin relaxation due to SOI shows $|\sigma_{x^{\prime\prime}}^{\uparrow\downarrow}|^2 = 1-\sin^2\theta_B\cos^2(\phi_B-\phi_d)$ dependencies on the orientation of the applied magnetic field and $\phi_d$ of the lateral separation with respect to the [100] direction.

\subsection{Intra-donor and intra-QD spin relaxation}

Besides the inter-donor-QD spin relaxation due to the hybridization of donor ground and QD states, there are also intra-donor and intra-QD spin relaxation mechanisms. Since spin relaxation $1/T_{1,donor}$ in a single donor, and spin relaxation $1/T_{1,QD}$ in a single QD in silicon have been studied intensively in the literature, \cite{roth1960,hasegawa1960,ramdas1981,tahan2002,khaetskii2001, golovach2004,tahan2014,huang_spin_2014} we discuss the intra-donor and intra-QD spin relaxation without deriving $1/T_{1,donor}$ and $1/T_{1,QD}$.

Suppose the effective Hamiltonian that causes spin-flip due to intra-donor and intra-QD mechanisms are denoted as $H_{1,donor}$ and $H_{1,QD}$; the spin relaxation in a single donor due to the finite matrix element $\bra{\uparrow} H_{1,donor} \ket{\downarrow}$ is $1/T_{1,donor}$; and the spin relaxation in a single QD due to the finite matrix element $\bra{\uparrow} H_{1,QD} \ket{\downarrow}$ is $1/T_{1,QD}$, then, the spin relaxation in a hybrid donor-QD system due to intra-donor and intra-QD mechanisms can be estimated accordingly.

Take the intra-QD mechanisms as an example, for the hybridized donor-QD state $\ket{\bar{n}}$, the spin flip matrix elements due to the intra-QD mechanism is given by
\beq
\bra{\bar{0}\uparrow} H_{1,QD} \ket{\bar{0}\downarrow}=\sum_{i,j\ne0} C_{\bar{0}i}^* C_{\bar{0}j} \bra{i\uparrow} H_{1,QD} \ket{j\downarrow}.
\eeq
If $t_{01},t_{02} \ll E_{VS}$, then, at least one of $C_{\bar{0}1}$ and $C_{\bar{0}2}$ will be much less than 1.
Note that, for a QD in silicon, the intra-QD mechanisms include the inter-valley spin relaxation and the intra-valley spin relaxation. \cite{yang2013, tahan2014, huang_spin_2014} For both mechanisms, we have that $\bra{1\uparrow} H_{1,QD} \ket{1\downarrow}= \bra{2\uparrow} H_{1,QD} \ket{2\downarrow}$. Therefore, we have
\beq
\bra{\bar{0}\uparrow} H_{1,QD} \ket{\bar{0}\downarrow}=(1-|C_{\bar{0}0}|^2) \bra{1\uparrow} H_{1,QD} \ket{1\downarrow}.
\eeq
\beq
1/T_{1,intra-QD}=(1-|C_{\bar{0}0}|^2)^2 1/T_{1,QD}.
\eeq
In this work,  both the inter-valley and the intra-valley mechanisms are included for the intra-QD spin relaxation. \cite{yang2013, huang_spin_2014}

For intra-donor mechanism, one can show that the spin relaxation is given by
\beq
1/T_{1,intra-donor}=|C_{\bar{0}0}|^4 1/T_{1,donor}.
\eeq
Therefore, the spin relaxation of a hybridized donor-QD state due to intra-QD (intra-donor) mechanisms is proportional to the spin relaxation in a single QD (donor) and proportional to the population of electron wavefunction in the QD (donor). In this work, we study only the phonon assisted spin relaxation for the intra-donor and intra-QD mechanisms.

\section{Results for inter-donor-QD spin relaxation}

In the following, we report the numerical results of spin relaxation as a function of the magnitude of the applied magnetic field and detuning of the donor and QD ground states. The dependence of spin relaxation with orientation of magnetic field is discussed in the Appendix.
Unless indicated, we choose the following parameters:
$t_{01} = t_{02} = 0.1$ meV,
$E_{VS}=0.3$ meV,
$g_{ani}$=0.001,
$\alpha_{so}=45$ m/s,
and $|d_{\parallel}|=2$ nm. For acoustic phonons in silicon, we choose
$v_1=5900$ m/s and $v_2=v_3=3750$ m/s for the speed of the different acoustic phonon branches, $\rho_c=2200$ kg/m$^3$ for the mass density,
$\Xi_d=5.0$ eV and
$\Xi_u=8.77$ eV for the dilation and shear deformation potential constants.
We choose $\theta_B=\pi/4$ and $\phi_B=0$ for the polar and azimuthal angles of the applied magnetic field. We choose $\phi_d=0$ for the angle of the in-plane QD shift $\vec{d}_{\parallel}$ relative to $[100]$.
For the intra-QD spin relaxation, we choose the orbital level splitting (confinement energy) $E_d=8$ meV, and $r^{-+}=1$ nm for the intra-QD relaxation. \cite{yang2013}
For the intra-donor spin relaxation, a bulk value $\Delta E_{1} = 12$ meV is assumed for the energy separation between the A state and E states of a phosphorus donor in silicon. \cite{kohn_theory_1955}

\subsection{$|B|$ dependence}

\begin{figure}
\includegraphics[scale=0.9]{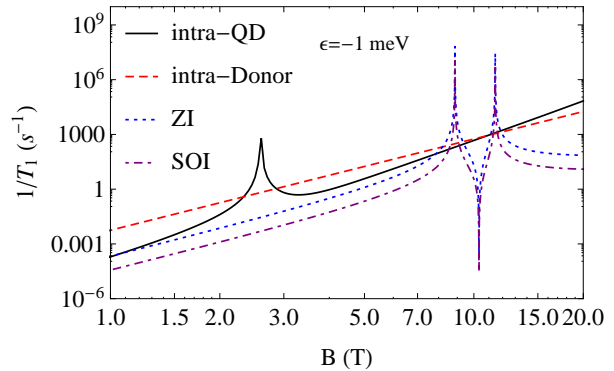}
\caption{Spin relaxation as a function of magnetic field when detuning $\epsilon=-1$ meV. We show spin relaxation due to intra-QD mechanism (black solid line), intra-donor mechanism (red dashed line), and inter-donor-QD mechanisms due to ZI (blue dotted line) and SOI (purple dot-dashed line).}\label{Fig_1}
\end{figure}

Figure \ref{Fig_1} shows the spin relaxation rate $1/T_1$ due to each mechanism as a function of the magnitude of the applied magnetic field when detuning $\epsilon= -1$ meV, although the donor ground state is lower than the QD states. The spin relaxation mechanisms include the intra-QD spin relaxation (black solid), intra-donor spin relaxation (red dashed), and inter-donor-QD spin relaxation due to ZI (blue dotted) and SOI (purple dot-dashed).
Because of the negative detuning, the ground state is mostly the donor ground state.
At low B-field, the spin relaxation is dominated by the intra-donor mechansim, and shows a $B^5$ dependence with B-field. As B-field increases, there are three spin relaxation peaks (hot-spots). One hot-spot is due to the intra-QD mechanism, where strong spin-valley hybridization occurs when the 1st excited orbital state $\ket{\bar{1}}$ crosses with the 2nd excited orbital states $\ket{\bar{2}}$ with opposite spin orientation; The two additional peaks are due to inter-donor-QD mechanisms (including ZI and SOI mechanisms), where strong hybridization occurs when the ground orbital state $\ket{\bar{0}}$ crosses with one of the two excited orbital states $\ket{\bar{r}}$ with opposite spin orientation. At higher B-field, the spin relaxation due to intra-QD mechanism becomes dominant, because of the stronger $B^7$ dependence with magnetic field.
For the inter-donor-QD mechanisms, there is also a spin relaxation dip (cool spot) between the two peaks due to destructive interference, whereas, it is masked by the intra-donor and intra-QD spin relaxation in this case.

\begin{figure}
\includegraphics[scale=0.64, bb = 0 0 340 243, clip=true]{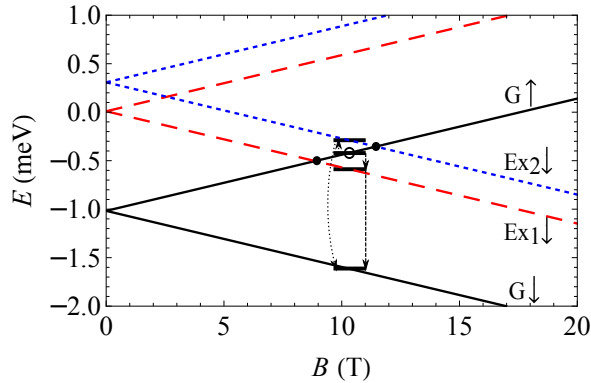}
\caption{Energy diagram as a function of magnetic field when detuning $\epsilon=-1$ meV. The ground orbital spin-up state crosses twice (marked as filled black dots) with two excited orbital spin-down states, which is responsible for the hot-spots of the inter-donor-QD spin relaxation in Fig. \ref{Fig_1}.
At the spin relaxation cool-spot (empty circle), a schematic diagram shows the two possible spin relaxation channels. 
}\label{Fig_1p}
\end{figure}

\begin{figure}
\includegraphics[scale=0.84, bb=0 0 102 141]{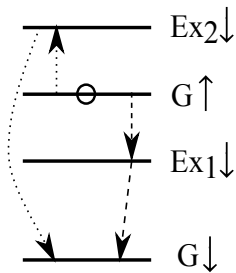}
\caption{A general schematic energy diagram at the spin relaxation cool-spot. There are two possible spin relaxation channels (dashed and dotted lines) indicated by lines with arrows, whose interference leads to the suppression of spin relaxation.
}\label{Fig_En_channels}
\end{figure}

Figure \ref{Fig_1p} shows the energy diagram as a function of magnetic field when detuning $\epsilon=-1$ meV. The Zeeman splitting for each orbital state increases with magnetic field.
The spin-up state of the 1st excited orbital crosses once 
with the spin-down state of the 2nd excited orbital, which is responsible for the intra-QD spin relaxation hot-spots in Fig. \ref{Fig_1}.
The ground orbital spin-up state crosses twice (marked as filled black dots) with two excited orbital spin-down states, which is responsible for the other two spin relaxation hot-spots in Fig. \ref{Fig_1}.
At the magnetic field marked by an empty circle, there is a spin relaxation cool-spot for the inter-donor-QD mechanisms, where a schematic diagram is shown for the energy levels and the spin relaxation channels.
Figure \ref{Fig_En_channels} shows a blow up of the energy diagram to highlight the two possible spin relaxation channels (dashed and dotted lines) that interfere to produce the spin relaxation cool-spot for the inter-donor-QD spin relaxation.


\begin{figure}
\includegraphics[scale=0.9]{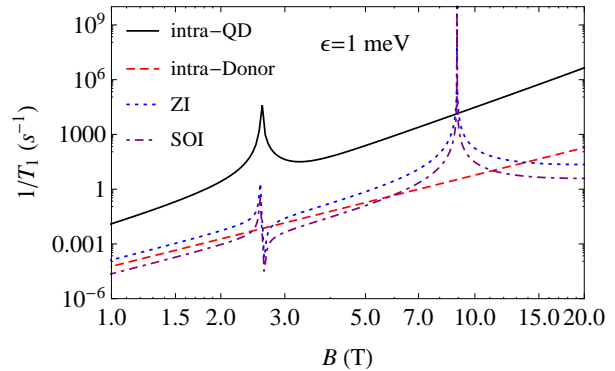}
\caption{Spin relaxation as a function of magnetic field when detuning $\epsilon=1$ meV.}\label{Fig_2}
\end{figure}


Figure \ref{Fig_2} shows the spin relaxation rate $1/T_1$ due to each mechanism as a function of the applied magnetic field for detuning $\epsilon= 1$ meV, where the QD states are lower than donor ground state.
For most B, the spin relaxation is dominated by the intra-QD mechanism.
The intra-QD mechanism results in a spin relaxation hot-spot due to spin-valley relaxation. 
The inter-donor-QD spin relaxation (include ZI and SOI mechanisms) results in two spin relaxation hot-spots, where one of hot-spot coincides with the hot-spot of intra-QD mechanism and the other hot-spot happens at higher B-field. Therefore, two spin relaxation hot-spots should be observable.
The two hot-spots happen when the ground orbital (mostly QD state $\ket{1}$) spin-up state crosses with spin-down states of the excited orbitals.
The spin relaxation cool-spot of the inter-donor-QD mechanism
is again due to destructive interference of relaxation channels. 
However, the cool-spot is still masked by the intra-QD spin relaxation.

In Figure \ref{Fig_2}, there is a dip very close to the first peak for the inter-donor-QD spin relaxation mechanisms. To understand the dip position, we study the condition of destructive interference. According to Eq. (\ref{eta_Z}), an interference dip appears when $\frac{E_{\bar{1}\bar{0}}|C_{\bar{0}0}^*C_{\bar{1}0}|^2}{E_{\bar{1}\bar{0}}^2 - E_Z^2} + \frac{E_{\bar{2}\bar{0}}|C_{\bar{0}0}^*C_{\bar{2}0}|^2}{E_{\bar{2}\bar{0}}^2 - E_Z^2}=0$. For $\epsilon= 1$ meV, we have $E_{\bar{1}\bar{0}}\approx E_{VS}=0.3$ meV, $E_{\bar{2}\bar{0}}\approx \epsilon=1$ meV, $|C_{\bar{0}0}|\approx \frac{t_{01}}{\epsilon}$, $|C_{\bar{1}0}|\approx \frac{t_{02}}{\epsilon}$, and $|C_{\bar{2}0}|\approx 1$. Therefore, the condition for destructive interference is ${E_Z} \approx \sqrt{E_{VS}^2 + \epsilon E_{VS}\frac{|C_{\bar{1}0}|^2}{C_{\bar{2}0}|^2}}$ $\approx  E_{VS} + t_{02}^2/(2\epsilon)$, and the distance between the peak and dip is $t_{02}^2/(2\epsilon)\sim 5$ $\mu$eV (or 43 mT), which explains the tiny separation of the peak and dip in this case.


\begin{figure}
\includegraphics[scale=0.9]{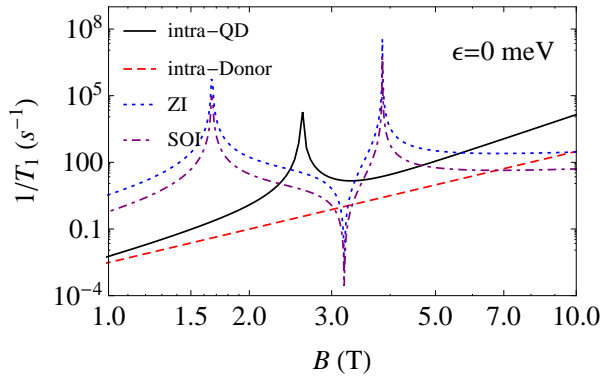}
\caption{Spin relaxation as a function of magnetic field when detuning $\epsilon=0$ meV.}\label{Fig_2pp} 
\end{figure}


Figure \ref{Fig_2pp} shows the same plot as Figure \ref{Fig_2}
when the detuning $\epsilon= 0$ meV, where the donor state is in resonance with the lower QD state and the hybridized orbital states are split by the tunneling coupling.
Due to the stronger orbital hybridization, the spin relaxation due to inter-donor-QD mechanisms dominate at low B-field, where the spin relaxation shows the $B^5$ dependence. Two spin relaxation peaks and one spin relaxation dip of the inter-donor-QD mechanisms appear similarly as before, but at lower B-fields. The hot-spot peak of the intra-QD relaxation shows up between the two peaks of the inter-donor-QD mechanisms. The magnitude of spin relaxation due to each mechanism indicates that three hot-spot will be observable, however, the cool-spot is again masked by the inter-QD mechanism.



\begin{figure}
\includegraphics[scale=0.9]{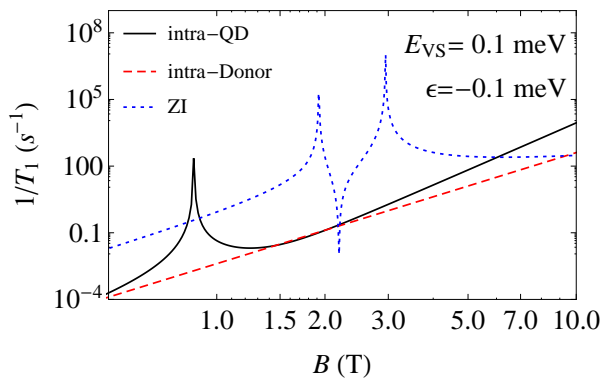}
\caption{Spin relaxation as a function of magnetic field when detuning $\epsilon=-0.1$ meV and $E_{VS}=0.1$ meV.}\label{Fig_2ppp}
\end{figure}

The previous discussion indicates that the cool-spot can be masked by the intra-QD mechanism, which makes it hard to observe in experiments. Moreover, there is in general no interference between the inter-donor-QD mechanisms and the intra-QD mechanism because of the phase difference of the effective spin-phonon interaction. For the cool-spot to be observed, one needs to suppress the intra-QD mechanism relative to the inter-donor-QD mechanism. In Figure \ref{Fig_2ppp}, we choose a small valley splitting $E_{VS}=0.1$ meV and small negative detuning $\epsilon=-0.1$ meV. Then, the spin relaxation hot-spot happens at lower magnetic field with smaller spin relaxation rate because of the reduced spin-valley relaxation \cite{huang_spin_2014}. The small negative detuning is chosen, so that the two-peaks of the inter-donor-QD mechanism happens right after the hot-spot of the intra-QD mechanism. In this case, the cool-spot of inter-donor-QD mechanism is no longer masked by the intra-QD mechanism, which makes it possible to be observed.

The spin relaxation cool-spot may also be masked by mechanisms other than spin relaxation due to intra-donor spin relaxation or intra-QD spin relaxation. For example, there is spin relaxation due to change of the effective mass and dielectric constants, spin relaxation due to phonon emission and Dresselhaus SOI, and the inter-donor-QD spin relaxation due to the coupling of valley states in the QD [second term on r.h.s. of Eq. (22)]. The electron spin relaxation rate due to change of the effective mass and dielectric constant was estimated to be around 500 minutes for phosphorus donors in silicon (1 T field), which is much less than the spin relaxation rate in our study \cite{pines_nuclear_1957}. The spin relaxation due to phonon emission and Dresselhaus SOI will show a coinciding cool-spot and hot-spot at the same magnitude of B. For the coupling of the valley states in QDs [second term on r.h.s. of Eq. (22)], it can in principle be much smaller than the ones we considered: 1) if the surface roughness is small, the inter-donor-QD spin relaxation due to the coupling of valley states should be smaller; 2) if the horizontal separation between donor and QD is increased, the spin relaxation we considered will be increased, without increasing the spin relaxation due to the coupling of valley states. We believe that, with proper system parameters, the spin relaxation cool-spot should in principle not be masked by other spin relaxation mechanisms.

One should note that the spin relaxation cool-spot of inter-donor-QD mechanism is due to the destructive interference of spin relaxation channels that have opposite sign of effective spin-phonon interaction, which happens when the ground orbital spin up state $\ket{\bar{0}\uparrow}$ is between the two excited orbital spin-down states $\ket{\bar{1}\downarrow}$ and $\ket{\bar{2}\downarrow}$. In the following, we discuss the effect of phonon coherence on the destructive interference.

For the destructive interference in this study, the phase coherence time of the channels is not essential because of the virtual nature of process of the interference.
From Eq. (\ref{Heff_ZI}) and Eq. (\ref{Heff_SO}), one can see that the effect of the electron-phonon interaction is independent with the factor $\eta_{Z}$, which characterizes the feature of interference. Destructive interference is due to the presence of multiple orbitals, and happens between the spin and orbital states. The interference is virtual. Thus, the time scale of an actual orbital flip is not important for an interference, because the virtual process does not require a real orbital excitation and orbital relaxation. However, the finite linewidth of phonon does play a role. The finite life time of phonon will break the energy conservation of the spin Zeeman energy and the energy of the emitted phonon, where the emitted phonon will not be exactly the spin Zeeman energy.
Thus, the finite phonon linewidth will broaden the feature of spin relaxation hot-spot and cool-spot.

Optical frequency measurement indicates that the phonon decays on the time scale of sub-nanoseconds at room temperature. However, for 10 GHz frequency phonon, which is emitted during the spin relaxation, the phonon coherence time is relatively long, it can be more than 10 ns. \cite{mante_thz_2015}
For a 10 ns coherence time of phonon at 10 GHz frequency, the broadening of phonon spectrum would correspond to a 0.05 Tesla difference in magnetic field, which will not destroy features of spin relaxation hot-spot and cool-spot in this study.
Since the interference involves multiple orbital states in both the donor and the QD, one would also require the phonon coherence length to be long enough to support the interference effect. 
Suppose that the phonon coherence length and the coherence time are limited by the same decoherence mechanism. Then, a phonon coherence time $\tau_{ph}=10$ ns would corresponds to coherence length $l_{ph}=v_\lambda \tau_{ph}\approx 10$ $\mu$s, which is much longer than the length scale of the coupled donor-QD system. Furthermore, the wave length of the emitted phonon is on the order of $2\pi v_\lambda/\omega_Z \approx 0.1$ $\mu$m, which is also much larger than the length scale of the coupled donor-QD system. In summary, the phonon coherence time and coherence length should be long enough to support the destructive interference of spin relaxation in a coupled donor-QD system at dilution refrigerator temperature.

\subsection{Detuning dependence}

\begin{figure}[t]
\includegraphics[scale=0.9]{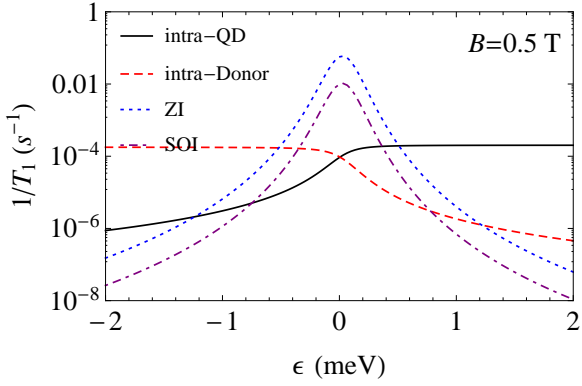}
\caption{Spin relaxation as a function of detuning when $B=0.5$ T. }\label{Fig_6}
\end{figure}

In this subsection, we study the detuning dependence of spin relaxation due to various mechanisms at different magnetic fields.

Figure \ref{Fig_6} shows the spin relaxation rate $1/T_1$ as a function of detuning between donor and interface states
when the applied magnetic field $B=0.5$ T ($\theta_B=\pi/4$ and $\phi_B=0$).
There is a broad transition of spin relaxation from negative to positive detuning for both spin relaxation mechanisms. There is no sharp peak because the orbital splitting is always larger than the spin Zeeman splitting, where the ground orbital spin up state does not cross with any excited orbital spin down state. Thus, hot-spot cannot happen in this case.

\begin{figure}[t]
\includegraphics[scale=0.9]{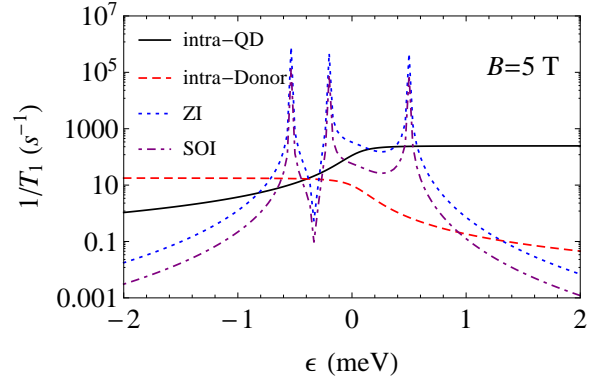}
\caption{Spin relaxation as a function of detuning when $B=5$ T.}\label{Fig_7}
\end{figure}


\begin{figure}[t]
\includegraphics[scale=0.63, bb=0 0 333 244]{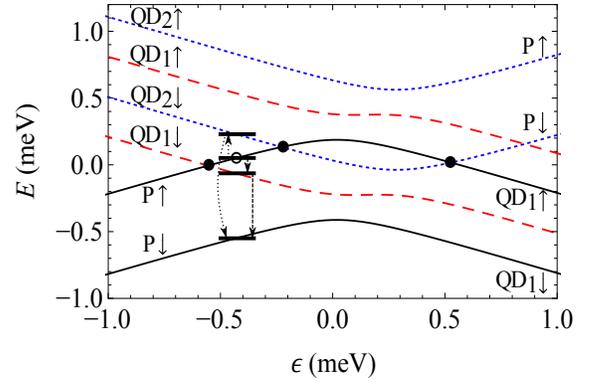}
\caption{Energy diagram as a function of detuning when $B=5$ T. The ground orbital spin-up state crosses three times with excited orbital spin-down states, which is responsible for the spin relaxation hot-spots in Fig. \ref{Fig_7}. A schematic diagram shows two possible spin relaxation channels (dashed and dotted lines) that can interference destructively. }\label{Fig_7p}
\end{figure}

Figure \ref{Fig_7} shows the spin relaxation rate $1/T_1$ as a function of the detuning when $B=5$ T. There are multiple peaks as the detuning goes from negative to positive values, which result from the multiple crossings of ground orbital spin up state and the excited orbital spin down states. There is also a cool-spot due to the interference of spin relaxation channels for the inter-donor-QD spin relaxation.
However, the cool-spot dip is masked by the intra-donor and intra-QD spin relaxation.

Figure \ref{Fig_7p} shows the corresponding energy diagram as a function of detuning when magnetic field $B=5$ T. The ground orbital spin-up state crosses three times with excited orbital spin-down states, which is responsible for the spin relaxation hot-spots in Fig. \ref{Fig_7}. At the spin relaxation cool-spot, a schematic diagram shows two possible spin relaxation channels (dashed and dotted lines), whose interference leads to the cancellation of the inter-donor-QD spin relaxation.

%
%
%


\begin{figure}[t]
\includegraphics[scale=0.9]{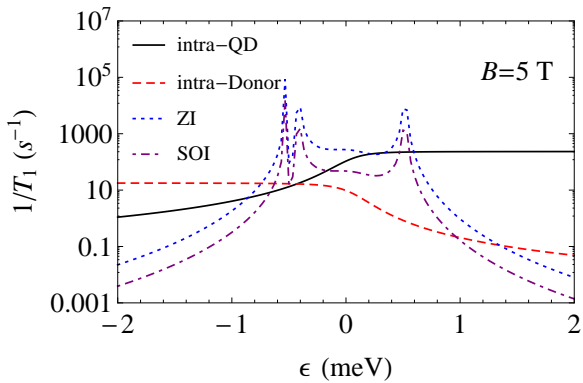}
\caption{Spin relaxation as a function of detuning when $B=5$ T and $E_{VS}=0.1$ meV.}\label{Fig_11}
\end{figure}

For the cool-spot to be observed, the intra-QD mechanism must be suppressed relative to the inter-donor-QD mechanism. In Figure \ref{Fig_11}, we choose a smaller valley splitting $E_{VS}=0.1$ meV. Then, the separation between two hot-spots is smaller, and the cool-spot of inter-donor-QD mechanism is no longer masked by the intra-QD or intra-donor spin relaxation. Note that the magnitude of intra-QD spin relaxation at $B=5$ T is dominated by the intra-valley spin relaxation, which is independent of the valley splitting $E_{VS}$.


Since detuning $\epsilon$ can be controlled electrically the spin relaxation hot-spots can be used to initialize spin states quickly and reduce the spin initialization error. The multiple hot-spots can also be used to study valley physics, such as valley splitting in QDs.
The qubit operation near a spin relaxation cool-spot could be used as a way to suppress the loss of quantum information (quantum coherence can still be limited by pure dephasing processes) during coherent transfer of spin information between donor atoms via interface states.

\section{Comparison with spin relaxation in a bulk donor in silicon}

Spin relaxation in a bulk P donor in silicon is dominated by ZI, which hybridizes the donor ground orbital spin-up state with the excited orbital spin down states. \cite{hasegawa1960,roth1960} The spin relaxation shows the $B^5$ dependence without peak or dip structures since the excited states of a bulk donor are at least 10 meV higher than the donor ground state, and the energy scale is much larger than the range of magnetic fields in experiment. The angular dependence of spin relaxation on the orientation of the applied magnetic field is given by $|\tensor{\Delta}_{XZ}^{(15)}|^2 + |\tensor{\Delta}_{XZ}^{(16)}|^2 + |\tensor{\Delta}_{YZ}^{(15)}|^2 + |\tensor{\Delta}_{YZ}^{(16)}|^2 \propto \sin^2\theta_B(4\cos^2\theta_B + \sin^2\theta_B\sin^22\phi_B)$  (see Appendix \ref{sec:bulkHz}, \ref{sec:bulkHep} and reference \cite{hasegawa1960}). The magnitude of the spin relaxation is on the order of $10^{-2}$ $s^{-1}$ for 1 T magnetic field along the [111] direction. \cite{hasegawa1960}

In comparison, the spin relaxation in a P donor coupled to interface states in silicon can be a result of ZI or SOI. Typically, the spin relaxation due to ZI is more important although it can be dominated by SOI in certain circumstances (with different B-field orientation and $d_\parallel$). The spin relaxation at low B-field shows the same $B^5$ dependence. When the detuning becomes small, there are peaks and dips in B-dependent spin relaxation due to the crossing of spin-orbit states and interference of spin relaxation channels. The angular dependence of spin relaxation on the orientation of the applied magnetic field is given by $|\tensor{\Delta}_{XZ}|^2 \propto \sin^2(2\theta_B)$.

Finally, we should mention that if the donor system is close to the metallic gates, where electrical noise such as Johnson noise could have a significant effect, then, the spin relaxation rate could show $B^3$ dependence with magnetic field because of the linear $\omega$ dependence of spectral density for Johnson noise; \cite{langsjoen_qubit_2012, poudel_relaxation_2013, huang_electron_2014, kha_micromagnets_2015, kolkowitz_probing_2015} Most of our analysis in this paper is also applicable when the donor is laterally coupled to a gate-defined QD, although the relative magnitude of spin relaxation rates due to ZI and SOI could be modified.

\section{Conclusion}

In conclusion, we have studied the spin relaxation of a donor coupled to QD-like interface states, we find both ZI with g-factor anisotropy and SOI can couple the donor ground state and QD states with opposite spin, and together with phonon emission will lead to inter-donor-QD spin relaxation.
We find that spin relaxation shows $B^5$ dependence at weak applied B-field.
Multiple spin relaxation hot-spots are found due to crossings of orbital states with opposite spin.
We find spin relaxation cool-spots due to the destructive interference of spin relaxation channels for the inter-donor-QD mechanism. While the cool-spot can be masked by the intra-donor and intra-QD spin relaxation, by fine tuning the system parameters, one can suppress the intra-donor and intra-QD spin relaxation, so that the spin relaxation cool-spot can be observable.
The qubit operations near spin relaxation hot-spots and cool-spots can be useful for the fast spin initialization and the preservation of quantum information during the transfer of spin qubit.
Finally, the orientation of the donor/QD geometry can be further used to suppress or enhance spin relaxation.

\section{Acknowledgement}

The authors thank J.M. Pomeroy (NIST) for useful discussions.
P.H. acknowledges the support by the Science, Technology and Innovation Commission of Shenzhen Municipality (No. ZDSYS20170303165926217, No. JCYJ20170412152620376)) and Guangdong Innovative and Entrepreneurial Research Team Program (Grant No. 2016ZT06D348) after joining in SUSTech.

\appendix

\section{Effective mass approximation and orbital basis}
\label{sec:EMA}

In this appendix, we review the effective mass theory for silicon and give explicitly the orbital eigenstates in terms of envelope function and Bloch function. We study cases when an electric potential is from a donor atom or a gate-defined QD. We also give the connection of the basis states used in main text and the eigenstates of a single donor and the eigenstates of a QD.

In the absence of the spin degree of freedom, the Schr\"{o}dinger equation is given by 
$(H_{Si}+V_e)\Phi_n(\vec{r})=E_n\Phi_n(\vec{r})$, where $H_{Si}$ is the unperturbed Hamiltonian of an electron near the conduction band minimum in a silicon crystal, $V_e(\vec{r})$ is the electrical potential from donor atom, interface and metallic gates, and $\Phi_n(\vec{r})$ is an eigenstate of the Hamiltonian. In a pure silicon crystal, the conduction band minima occur near the X symmetry points with six-fold degeneracy, normally referred to as valley degeneracy. In the presence of the potential $V_e(\vec{r})$, the valley degeneracy is lifted. The wavefunction can be written as
\beq
\Phi_n(\vec{r})=\sum_{j=1}^6 a_n^{(j)} F_n^{(j)}(\vec{r})\psi_j(\vec{r}), \label{Phin}
\eeq
where $j$ is the valley index, $\psi_j(\vec{r})=e^{i\vec{k}_j\cdot\vec{r}} u_j(\vec{r})$ is a Bloch function, $\vec{k}_j$ is one of the six minima $\pm k_0 \hat{x}$, $\pm k_0 \hat{y}$ or $\pm k_0 \hat{z}$ ($k_0=0.85\cdot2\pi/a_0$ and $a_0$ is the lattice constant), and $u_j(\vec{r})$ is the periodic function (Bloch's theorem). $F_n^{(j)}(\vec{r})$ is the slowly varying envelope function, and $a_n^{(j)}$ is the probability amplitude of $j$-th valley state due to valley coupling in the presence of a sharp potential.

Within the effective mass approximation, $F_n^{(j)}(\vec{r})$ satisfies a Schr\"{o}dinger-like equation for the $j$-th valley. For example, $F_n^{(\pm z)} (\vec{r})$ satisfies
\beqa
\bigg[-\frac{\hbar^2}{2m_l}\frac{\partial^2}{\partial z^2} - \frac{\hbar^2}{2m_t}\left(\frac{\partial^2}{\partial x^2} + \frac{\partial^2}{\partial y^2} \right) + V_e(\vec{r}) \bigg] F_n^{(\pm z)}(\vec{r}) \nonumber\\
=(E_n^{(\pm z)}-E_{k_0}^{(\Delta)}) F_n^{(\pm z)} (\vec{r}), \nonumber
\eeqa
where $m_l$ and $m_t$ are the longitudinal and transverse mass, $E_{k_0}^{(\Delta)}$ is the energy of band minimum at $\vec{k}=\pm k_0\hat{z}$, $E_n^{(\pm z)}$ is the eigenenergy. Similar equations can be given for the envelope functions $F_n^{(\pm x)} (\vec{r})$ and $F_n^{(\pm y)} (\vec{r})$ at $\vec{k}=\pm k_0\hat{x}$ and $\pm k_0\hat{y}$ minima.
A sharp potential can further couple different valley states of $F_n^{(j)}(\vec{r})\psi_j(\vec{r})$. By solving the coupled multi-valley Schr\"{o}dinger equation, the eigenstates $\Phi_n(\vec{r})$ in Eq. (\ref{Phin}) can be obtained, and valley degeneracy can be lifted due to the broken of symmetry in presence of potential $V_e(\vec{r})$.


According to the symmetry, when $V_e(\vec{r})$ is the potential of a single P donor in silicon, the coefficients of the six lowest eigenstates as given by Kohn and Luttinger, and modified to be orthogonal are,\cite{kohn_theory_1955,wilson_electron_1961}
\beqa
a_{P,1}&=&1/\sqrt{6}[1,1,1,1,1,1], \nonumber\\
a_{P,2}&=&1/\sqrt{2}[1,-1,0,0,0,0], \nonumber\\
a_{P,3}&=&1/\sqrt{2}[0,0,1,-1,0,0], \nonumber\\
a_{P,4}&=&1/\sqrt{2}[0,0,0,0,1,-1], \nonumber\\
a_{P,5}&=&1/\sqrt{12}[-1,-1,-1,-1,2,2], \nonumber\\
a_{P,6}&=&1/2[1,1,-1,-1,0,0], \nonumber
\eeqa
where the basis used is $[F_P^{(x)}\psi_{x}$, $F_P^{(\bar{x})}\psi_{\bar{x}}$, $F_P^{(y)}\psi_{y}$, $F_P^{(\bar{y})}\psi_{\bar{y}}$, $F_P^{(z)}\psi_{z}$, $F_P^{(\bar{z})}\psi_{\bar{z}}]$ and $F_P^{(j)}(\vec{r}-\vec{R}_{P})$ is the ground state envelope function that satisfies the single-valley Schr\"{o}dinger-like equation with $V_e(\vec{r})$ a single donor potential.

When $V_e(\vec{r})$ is the potential of a single QD at a $z$ interface, the coefficients of the two lowest eigenstates are
\beqa
a_{QD,1}&=&1/\sqrt{2}[0,0,0,0,e^{i\phi_v},e^{-i\phi_v}], \nonumber\\
a_{QD,2}&=&1/\sqrt{2}[0,0,0,0,e^{i\phi_v},-e^{-i\phi_v}], \nonumber
\eeqa
where the basis is $[F_{QD}^{x}\psi_{(x)}$, $F_{QD}^{(\bar{x})}\psi_{\bar{x}}$, $F_{QD}^{(y)}\psi_{y}$, $F_{QD}^{(\bar{y})}\psi_{\bar{y}}$, $F_{QD}^{(z)}\psi_{z}$, $F_{QD}^{(\bar{z})}\psi_{\bar{z}}]$ and $F_{QD}^{(j)}(\vec{r}-\vec{R}_{QD})$ is the ground state envelope function that satisfies the single-valley Schr\"{o}dinger-like equation with $V_e(\vec{r})$ the QD potential, $\phi_v$ is the valley phase difference between $+z$ and $-z$ valleys.\cite{zimmerman_valley_2017}

In the main text, we use basis state $\ket{n}$ of a donor ground state and two lowest QD states, which can be expressed in terms of envelope and Bloch functions
\beq
\langle r\ket{n}=\sum_{j=1}^6 \alpha_n^{(j)} F_n^{(j)}(\vec{r})\psi_j(\vec{r}),
\eeq
where the coefficients $\alpha_n$ is related to $a_P$ and $a_{QD}$:
$\alpha_0=a_{P,1}$, $\alpha_1=a_{QD,1}$, $\alpha_2=a_{QD,2}$; The envelope functions $F_0^{(j)}$ and $F_1^{(j)} = F_2^{(j)}$ are the orthornormalized states of $F_P^{(j)}$ and $F_{QD}^{(j)}$. 

\section{ZI when $\tensor{g}^{(j)}$ is different for donor and QD}
\label{sec:Hz}

In this appendix, a general form of the ZI is studied when the g-factor is different for donor and QD. Based on the general form of ZI, we study the spin-orbit hybridization due to the ZI. We find that the hybridization exhibits a similar form as in the main text. The only difference is that the hybridization is proportional to the g-factor anisotropy $g_{QD,ani}$ in a QD instead of $g_{ani}$.

In the presence of an external magnetic field, the ZI is given by
\beq
H_Z=\frac{1}{2}\mu_B \sum_{nj} \ket{F_n^{(j)}\psi_j}\bra{F_n^{(j)}\psi_j} \vec{\sigma}\cdot \tensor{g}^{(n,j)} \cdot \vec{B},
\eeq
where $\tensor{g}^{(n,j)}$ are the g-factor tensor when an electron is in $j$-th valley with envelope function $F_n^{(j)}$.
g-factor tensor $\tensor{g}^{(n,j)}$ is anisotropic (which is also a result of SOI that hybridizes spin and electronic bands),
\beq
\tensor{g}^{(n,j)} =g_{n,\parallel} \tensor{U}^{(j)} + g_{n,\perp} \left(\tensor{1}-\tensor{U}^{(j)}\right),
\eeq
where $j$ is the index for six valley states in silicon, $g_{n,\parallel}$ ($g_{n,\perp}$) is the g-factors along (perpendicular to) the valley ellipsoid for an electron in the $n$-th orbital state. $\tensor{U}^{(j)}$ is the projection operator in the 3-dimensional coordinate space, for example,
when $j=z$, we have
\beq
\tensor{U}^{(z)}=
\left[
\begin{array}{ccc}
0 & 0 & 0 \\
0 & 0 & 0 \\
0 & 0 & 1
\end{array}
\right],
\tensor{1}=
\left[
\begin{array}{ccc}
1 & 0 & 0 \\
0 & 1 & 0 \\
0 & 0 & 1
\end{array}
\right].
\eeq
Alternatively, g-factor tensor can be written as
\beqa
\tensor{g}^{(n,j)}
&=&g_{n,\perp} \tensor{1} + g_{n,ani} \tensor{U}^{(j)},
\eeqa
where $g_{n,\perp}=g_{n,avg}-1/3 g_{n,ani}$, $g_{n,avg}=(g_{n,\parallel}+2g_{n,\perp})/3$, and $g_{n,ani}=g_{n,\parallel}-g_{n,\perp}$. Thus, one can see that the g-factor anisotropy $g_{n,ani}$ causes the hybridization of spin and valley states within the donor eigenstates. This is known for an electron binding to a donor atom, where g-factor anisotropy will couple the ground state and excited donor orbital states with opposite spin orientation.

We express the ZI in the basis of donor ground state $\ket{0}$ and QD ground states $\ket{1}$ and $\ket{2}$:
\beq
H_{Z,nn^\prime}
= \frac{1}{2}\mu_B \vec{\sigma}\cdot  \tensor{g}^{(nn^\prime)} \cdot \vec{B},
\eeq
where
\beq
\tensor{g}^{(nn^\prime)}=\left(g_{\perp} \tensor{1} + g_{ani} \tensor{D}^{(nn^\prime)}\right)\bra{F_n^{(j)}} \left.F_{n^\prime}^{(j)} \right\rangle,
\eeq
\beqa
\tensor{D}^{(nn^\prime)}=\sum_{j} \alpha_{n j}\alpha_{n^\prime j} \tensor{U}^{(j)}.
\eeqa
If $n\ne n^\prime$, one can find that $H_{Z,nn^\prime}=0$. Therefore, we have
\beq
\tensor{g}^{(nn^\prime)}=\left(g_{n,\perp} \tensor{1} + g_{n,ani} \tensor{D}^{(nn)}\right) \delta_{nn^\prime}.
\eeq
If $n=n^\prime=0$, then,
$\tensor{D}^{(00)}=\frac{1}{6}\sum_{j} \tensor{U}^{(j)}=\frac{1}{3}\tensor{1}.
$
Thus,
\beq
\tensor{g}^{(00)}=g_{p,\perp} \tensor{1} + \frac{1}{3} g_{P,ani} \tensor{1}=g_{avg} \tensor{1},
\eeq
\beqa
H_{Z,00}&=& \frac{1}{2}g_{P,avg}\mu_B \vec{\sigma}\cdot \vec{B},
\eeqa
where $g_{P,avg}=g_{P,\perp}+ g_{P,ani}/3$.
If $n=n^\prime=1$ or 2, then,
$
\tensor{D}^{(11)}=\tensor{D}^{(22)}=\tensor{U}^{(z)}.
$
Thus, 
\beq
\tensor{g}^{(11)}=\tensor{g}^{(22)}=g_{QD,\perp} \tensor{1} + g_{QD,ani} \tensor{U}^{(z)},
\eeq
\beqa
H_{Z,11} 
&=&\frac{1}{2}\mu_B \left(g_{QD,\perp}\vec{\sigma}\cdot \vec{B} + g_{QD,ani} \sigma_zB_z\right).
\eeqa
Then, we can express ZI in the basis of orbital eigenstates $\ket{\bar{n}}$. Since
\beq
\ket{\bar{n}}=\sum_{n} C_{\bar{n}n}\ket{{n}} = \sum_{n j} C_{\bar{n}n} \alpha_{n j}\ket{F_n^{(j)} \psi_{j}},
\eeq
we have
\beqa
(H_{Z})_{\bar{0}\bar{n}}
&=& \sum_{n}  C_{\bar{0}n}^*C_{\bar{n}n} H_{Z,nn^\prime},
\eeqa
Therefore, the ZI in the basis of orbital eigenstates is 
\beqa
(H_Z)_{\bar{0}\bar{n}} &=& \frac{1}{2}\mu_B \vec{\sigma}\cdot  \tensor{G}^{(\bar{0}\bar{n})} \cdot \vec{B},
\eeqa
\beqa
\tensor{G}^{(\bar{0}\bar{n})} 
= |C_{\bar{0}0}|^2\tensor{g}^{(00)} \delta_{\bar{0}\bar{n}}+ C_{\bar{0}0}^*C_{\bar{n}0} (\tensor{g}^{(00)} - \tensor{g}^{(11)} ), 
\eeqa
\beqa
\tensor{g}^{(00)} - \tensor{g}^{(11)} = (g_{P,avg}-g_{QD,\perp}) \tensor{1} - g_{QD,ani} \tensor{U}^{(z)},
\eeqa
where the orthogonal relations $\sum_n C_{\bar{0}n}^*C_{\bar{n}n}=\delta_{\bar{0}\bar{n}}$ have been employed.
%
Therefore,
\beqa
\tensor{G}^{\bar{0}\bar{n}} &=& 
\left[
\begin{array}{ccc}
\tensor{G}^{\bar{0}\bar{n}}_{xx} &  &   \\
    & \tensor{G}^{\bar{0}\bar{n}}_{xx} &  \\
    &             & \tensor{G}^{\bar{0}\bar{n}}_{zz}
\end{array}\right],
\eeqa
\beqa
\tensor{G}^{\bar{0}\bar{n}}_{xx}&=&g_{QD,\perp} \delta_{\bar{0}\bar{n}} + C_{\bar{0}0}^*C_{\bar{n}0} (g_{P,avg} - g_{QD,\perp}),\\
\tensor{G}^{\bar{0}\bar{n}}_{zz}&=&g_{QD,\parallel} \delta_{\bar{0}\bar{n}} + C_{\bar{0}0}^*C_{\bar{n}0}(g_{P,avg} - g_{QD,\parallel}),
\eeqa
which is consistent with the results in the main text when the g-factor tensor is the same for donor and QD. 
Therefore, the ZI can be expressed as
\beq
(H_Z)_{\bar{0}\bar{n}}= \frac{1}{2} \mu_B B  \sum_{\xi=x,y,z} g_{\xi}^{\bar{0}\bar{n}} \sigma_{\xi},
\eeq
where $g_{\xi}^{\bar{0}\bar{n}} =  G_{\xi\xi}^{\bar{0}\bar{n}}\hat{b}_{\xi}$, and $\hat{b}=[\hat{b}_x$, $\hat{b}_y$, $\hat{b}_z]$ = $[\sin\theta_B\cos\phi_B$, $\sin\theta_B\sin\phi_B$, $\cos\theta_B]$ is a unit vector along the applied magnetic field.

In order to find the effective spin flip matrix element, we need to express the interaction terms in a new $(X,Y,Z)$ coordinate system, where the $Z$ axis is along the spin quantization axis determined by $(H_{Z})_{\bar{0}\bar{0}}$. When the $Z$ axis is along spin quantization axis, $(H_{Z})_{\bar{0}\bar{0}}$ is diagonalized. Note that, because of the anisotropy of $ G_{\xi\xi}^{\bar{0}\bar{0}}$, the spin quantization axis given by $(g_{x}^{\bar{0}\bar{0}}, g_{y}^{\bar{0}\bar{0}}, g_{z}^{\bar{0}\bar{0}})$ is slightly different from the direction of B-field.
An Euler rotation can be done to rotate $(x,y,z)$ to $(X,Y,Z)$ coordinate system.
In the new coordinate system, $(H_{Z})_{\bar{0}\bar{r}}$ is in general expressed as
\beqa
(H_{Z})_{\bar{0}\bar{r}} &\equiv&  \frac{1}{2} \mu_B B  \sum_{\xi=X,Y,Z} g_{\xi}^{\bar{0}\bar{r}} \sigma_{\xi}.
\eeqa
where $g_{X}^{\bar{0}\bar{r}}$ and $g_{Y}^{\bar{0}\bar{r}}$ are relevant to spin-orbit hybridization.

In general, Euler rotation angles can be obtained to transform from $(x,y,z)$ to $(X,Y,Z)$ coordinate system. However, since $g_x^{\bar{0}\bar{0}}/g_y^{\bar{0}\bar{0}}=g_x^{\bar{0}\bar{r}}/g_y^{\bar{0}\bar{r}}=\hat{b}_x/\hat{b}_y$, the calculation can be simplified. We first rotate $(x,y,z)$ to $(x^\prime,y^\prime,z)$, where $x^\prime$ is along the projection of $\vec{B}$ on $(x,y)$ plane. Consequently, $[B_{x^\prime},B_{y^\prime},B_z]$ = $B[\sin\theta_B,0,\cos\theta_B]$, $[g_{x^\prime}^{0n}, g_{y^\prime}^{0n}, g_{z}^{0n}]$ = $[\tensor{G}^{\bar{0}\bar{n}}_{xx}\sin\theta_B, 0, \tensor{G}^{\bar{0}\bar{n}}_{zz}\cos\theta_B]$, and $(H_{Z})_{\bar{0}\bar{0}}$ = $E_Z/2[\sin\theta_s,0,\cos\theta_s]\cdot \vec{\sigma}$, where $E_Z = g_{eff} \mu_B B$, $g_{eff} =\sqrt{(\tensor{G}^{\bar{0}\bar{0}}_{xx}\sin\theta_B)^2+ (\tensor{G}^{\bar{0}\bar{0}}_{zz}\cos\theta_B)^2}$, and $\tan\theta_s=(\tensor{G}^{\bar{0}\bar{0}}_{xx}/\tensor{G}^{\bar{0}\bar{0}}_{zz})\tan\theta_B$.
Then, a rotation $R_y(-\theta_s)$ in $(x^\prime, z)$ plane from $(x^\prime, z)$ to $(X, Z)$
will diagonalize $(H_{Z})_{\bar{0}\bar{0}}$,
\beq
(H_{Z})_{\bar{0}\bar{0}} = \frac{1}{2} E_Z \sigma_{Z}.
\eeq
Correspondingly, the spin-orbit hybridization term due to ZI is
\beqa
(H_{Z})_{\bar{0}\bar{r}} &\equiv&  \frac{1}{2} \mu_B B  (g_{X}^{\bar{0}\bar{r}} \sigma_{X} + g_{Z}^{\bar{0}\bar{r}} \sigma_{Z}),
\eeqa
where
\beqa
g_X^{\bar{0}\bar{r}} &=& \tensor{G}^{\bar{0}\bar{r}}_{xx}\sin\theta_B\cos\theta_s - \tensor{G}^{\bar{0}\bar{r}}_{zz}\cos\theta_B\sin\theta_s, \\
g_Z^{\bar{0}\bar{r}} &=& \tensor{G}^{\bar{0}\bar{r}}_{xx}\sin\theta_B\sin\theta_s + \tensor{G}^{\bar{0}\bar{r}}_{zz}\cos\theta_B\cos\theta_s.
\eeqa
Note that, for the spin flip process, only the term $\frac{1}{2} \mu_B B  g_{X}^{\bar{0}\bar{r}}\sigma_X$ is relevant, where
\beqa
g_X^{\bar{0}\bar{r}} 
&=& \frac{1}{2} (\tensor{G}^{\bar{0}\bar{r}}_{xx}- \tensor{G}^{\bar{0}\bar{r}}_{zz})\sin(\theta_B+\theta_s) \nonumber\\&+&
\frac{1}{2} (\tensor{G}^{\bar{0}\bar{r}}_{xx} + \tensor{G}^{\bar{0}\bar{r}}_{zz})\sin(\theta_B-\theta_s) \nonumber\\
&\approx&\frac{1}{2} (\tensor{G}^{\bar{0}\bar{r}}_{xx} - \tensor{G}^{\bar{0}\bar{r}}_{zz}) \sin(2\theta_B) \nonumber\\
&\equiv& \frac{1}{2}g_{QD,ani}C_{\bar{0}0}^*C_{\bar{r}0} \sin(2\theta_B).
\eeqa
Thus, the only difference in comparison with the results in the main text is that the hybridization is proportional to $g_{QD,ani}$ in stead of $g_{ani}$.

\section{Electron-phonon interaction}  
\label{sec:ep}

In this appendix, we derive the Hamiltonian $H_{EP}$ and obtain the explicit coefficients of $\Pi_{\vec{q}\lambda}^{(j)}$ for the electron-phonon interaction in silicon.

An electron in a semiconductor conduction band interacts with phonons (lattice vibration) when the band energy shifts under elastic strain. This electron-phonon interaction is
\beq
H_{EP}=\sum_{\alpha\beta}\tensor{\Xi}_{\alpha\beta}\tensor{\varepsilon}_{\alpha\beta}, 
\eeq
where $\tensor{\Xi}_{\alpha\beta}$ is the deformation potential tensor and $\tensor{\varepsilon}_{\alpha\beta}$ is the strain tensor ($\alpha,\beta=x,y,z$).

In silicon, there are six valley states, and the electron-phonon deformation potential (without Umklapp processes) can be expressed as
\beq
\tensor{\Xi} = \sum_j P_j\tensor{\Xi}^{(j)},
\eeq
\beq
\tensor\Xi^{(j)}=\Xi_{d}\tensor{1}+ \Xi_{u}\tensor{U}^{(j)},
\eeq
where $P_j=\ket{\psi_j}\bra{\psi_j}$ is the projection operator that selects the $j$-th valley, $\Xi_d$ and $\Xi_u$ are the dilation and uniaxial shear deformation potential constants.

\begin{table}[t]
  \centering
\[
\begin{array}{c|c|c|c}
  \hline\hline
   \hat{e}_\alpha^{(\vec{q}\lambda)}   & \lambda=l & \lambda=t_1 & \lambda=t_2 \\ \hline
  \alpha=x & \sin\vartheta\cos\varphi & \cos\vartheta\cos\varphi & -\sin\varphi  \\ \hline
  \alpha=y & \sin\vartheta\sin\varphi & \cos\vartheta\sin\varphi & \cos\varphi \\ \hline
  \alpha=z & \cos\vartheta         & -\sin\vartheta & 0         \\
  \hline\hline
\end{array}
\]
  \caption{Polarization unit vector components for different phonon branches (including LA, TA1 and TA2).}\label{polarization components}
\end{table}

The strain tensor $\tensor{\varepsilon}_{\alpha\beta}$ is ($\alpha,\beta=x,y,z$)
\beq
\tensor{\varepsilon}_{\alpha\beta}=\frac{1}{2}\left(\frac{\partial{u}_{\alpha}}{\partial r_{\beta}} + \frac{\partial u_{\beta}}{\partial r_{\alpha}}\right),
\eeq
where $r_{\alpha}$, $r_{\beta}$ are the coordinates and $u_\alpha$, $u_\beta$ are the lattice displacement under strain. The only phonons involved are acoustic phonons, one longitudinal (l) and two transverse ($t_1$, $t_2$).
The phonon displacement is given by 
\begin{equation}
\vec{u}_{\vec{q}\lambda}(\vec{r})=\sqrt{\hbar/2\rho_c\omega_{q\lambda}}\hat{e}^{(\vec{q}\lambda)} e^{i\vec{q}\cdot\vec{r}}(b_{-\vec{q}\lambda}^\dag + b_{\vec{q}\lambda})
\end{equation}
where $b_{\vec{q}\lambda}^\dag$ ($b_{\vec{q}\lambda}$) is the creation (annihilate) operator of a phonon with wave vector $\vec{q}$ and branch-index $\lambda$, $\hat{e}_{\vec{q}\lambda}$ is the polarization unit vector (see Table \ref{polarization components}), $\rho_c$ is the sample density (volume is set to unity here). Therefore, the strain tensor is
\[
\tensor{\varepsilon}_{\alpha\beta}=\sum_{\vec{q}\lambda}\frac{i}{2}\sqrt{\frac{\hbar q}{2\rho_cv_{q\lambda}}}(\hat{e}_\alpha^{(\vec{q}\lambda)} \hat{q}_{\beta}+\hat{e}_\beta^{(\vec{q}\lambda)} \hat{q}_{\alpha})e^{i\vec{q}\cdot\vec{r}} (b_{-\vec{q}\lambda}^\dag + b_{\vec{q}\lambda}),
\]
where $\hat{q}=\vec{q}/|q|$ is the unit vector along $\vec{q}$ and $v_{q\lambda}$ is the velocity of the corresponding phonon mode.

Therefore, electron-phonon interaction in silicon is
\begin{equation}
H_{EP}=\sum_j P_j \sum_{\vec{q}\lambda}e^{i\vec{q}\cdot\vec{r}} M_{\vec{q}\lambda}^{(j)} (b_{-\vec{q}\lambda}^\dag + b_{\vec{q}\lambda}), 
\end{equation}
\beq
M_{\vec{q}\lambda}^{(j)} =i\sqrt{\hbar q/2\rho_c v_{\lambda}} \Pi_{\vec{q}\lambda}^{(j)}, 
\eeq
\beqa
{\Pi}_{\vec{q}\lambda}^{(j)}
&=& \hat{e}^{(\vec{q}\lambda)} \cdot \tensor{\Xi}^{(j)} \cdot \hat{q},
\eeqa
where the coefficient ${\Pi}_{\vec{q}\lambda}^{(j)}$ determines the strength of electron-phonon interaction.

When the electron is in $(0,0,\pm k_0)$ valleys, we have
\beq
\Pi_{\vec{q}\lambda}^{(z)} = \Xi_d\hat{e}_x^{(\vec{q}\lambda)} \hat{q}_{x}+\Xi_d\hat{e}_y^{(\vec{q}\lambda)} \hat{q}_{y}+(\Xi_d+\Xi_u)\hat{e}_z^{(\vec{q}\lambda)} \hat{q}_{z},
\eeq
Note that $\hat{q}=\hat{e}^{(\vec{q}l)}$=$[\sin\vartheta\cos\varphi$, $\sin\vartheta\sin\varphi$, $\cos\vartheta]$.
Thus, for an electron in the $\pm$z valleys, we have $\Pi_{\vec{q},l}^{(z)}=\Xi_d+\Xi_u\cos^2\vartheta$, $\Pi_{\vec{q},t_1}^{(z)}=-\Xi_u\sin\vartheta\cos\vartheta$, and $\Pi_{\vec{q},t_2}^{(z)}=0$.
For an electron in an arbitrary valley state, the coefficient $\Pi_{\vec{q}\lambda}^{(j)}$ can be obtained as summarized in Table (\ref{tableXi}). 
\begin{widetext}

\begin{table}[ht]
\[
\begin{array}{c|c|c|c}
  \hline\hline
  \Pi_{\vec{q}\lambda}^{(j)} & \lambda=l & \lambda=t_1 & \lambda=t_2 \\ \hline
  j=\pm x\mathrm{-valley} &\Xi_d+\Xi_u\sin^2\vartheta\cos^2\varphi & \Xi_u\sin\vartheta\cos\vartheta\cos^2\varphi & - \Xi_u\sin\vartheta\cos\varphi\sin\varphi  \\ \hline
  j=\pm y\mathrm{-valley}  &\Xi_d+\Xi_u\sin^2\vartheta\sin^2\varphi & \Xi_u\sin\vartheta\cos\vartheta\sin^2\varphi  & \Xi_u\sin\vartheta\cos\varphi\sin\varphi \\ \hline
  j=\pm z\mathrm{-valley} & \Xi_d+\Xi_u\cos^2\vartheta        & - \Xi_u\sin\vartheta\cos\vartheta  & 0         \\  \hline\hline
\end{array}
\]
  \caption{Coefficients $\Pi_{\vec{q}\lambda}^{(j)}$ for an electron in $j$-th valley interacting with the $\lambda$-th branch of phonons.}\label{tableXi}
\end{table}
\end{widetext}

Based on the coefficients in Table (\ref{tableXi}), the averaged value for donor ground and QD ground states
\beq
{\Pi}_{\vec{q}\lambda,nn} =  \bra{n} \sum_j P_j \Pi_{\vec{q}\lambda}^{(j)} \ket{n} =\sum_j \alpha_{n}^{(j)}\alpha_{n}^{(j)} {\Pi}_{\vec{q}\lambda}^{(j)},
\eeq
can be obtained as shown in table \ref{tableXinn}. Then, 
\beqa
{\Pi}_{\vec{q}\lambda}^\prime
&=& {\Pi}_{\vec{q}\lambda,00} - {\Pi}_{\vec{q}\lambda,11} = \Xi_{u}\hat{e}^{(\vec{q}\lambda)} \cdot \tensor{\Delta} \cdot \hat{q} ,
\eeqa
is also obtained (see Table \ref{tableXinn}).


\begin{table}[h]
\[
\begin{array}{c|c|c|c}
  \hline\hline
{\Pi}_{\vec{q}\lambda,nn} &\lambda=l &\lambda=t_1 & \lambda=t_2 \\ \hline
  n=0 \texttt{ (P)} &\Xi_d +\Xi_u/3 & 0 & 0 \\ \hline
  n=1,2 \texttt{ (QD)} & \Xi_d+\Xi_u\cos^2\vartheta         & -\Xi_u\sin\vartheta\cos\vartheta & 0         \\  \hline\hline
{\Pi}_{\vec{q}\lambda}^\prime & \Xi_u(1/3-\cos^2\vartheta)        & \Xi_u\sin\vartheta\cos\vartheta & 0         \\  \hline\hline
\end{array}
\]
  \caption{Averaged coefficients $\Pi_{\vec{q}\lambda,nn}$ for donor ground and QD ground states as well as the coefficient ${\Pi}_{\vec{q}\lambda}^\prime = {\Pi}_{\vec{q}\lambda,00}-{\Pi}_{\vec{q}\lambda,11}$ for the $\lambda$-th branch of phonons.}\label{tableXinn}
\end{table}

\section{Commutation relation}
\label{sec:commutation}

In this appendix we study the commutation property of $[x,H_O]$ in silicon for the evaluation of $(p_x)_{\bar{0}\bar{r}}$ within the multivalley effective mass approximation. To evaluate the matrix element $(p_x)_{\bar{0}\bar{r}}$, it is convenient if we have $[x,H_O]=i\hbar p_x/m^*$, which is valid in the case of single valley physics. Since the effective mass is different for different valleys, we need to re-derive a new commutation relation.

We consider the orbital Hamiltonian
\beqa
H_O &=& H_K + V = \sum_{j}\sum_i \frac{(p_i^{(j)})^2}{2m_i^{(j)}}\ket{j}\bra{j} + V(\vec{r}) \nonumber\\
&=& \sum_{j} \left[\sum_i \frac{(p_i^{(j)})^2}{2m_i^{(j)}}  + V_{jj} \right]\ket{j}\bra{j}
+ \sum_{jj^\prime}V_{jj^\prime} \ket{j}\bra{j^\prime}, \nonumber
\eeqa
where $i=x$, $y$, $z$, and $j$ is the valley index. In the kinetic term $H_K$, we have $m^{(\pm z)}=(m_x^{(\pm z)}, m_y^{(\pm z)}, m_z^{(\pm z)})=(m_t,m_t,m_l)$, where $m_t$ and $m_l$ are the transverse and longitudinal effective mass in silicon; similarly, we have $m^{(\pm x)}=(m_l,m_t,m_t)$, and $m^{(\pm y)}=(m_t,m_l,m_t)$. We have assumed that the kinetic term $H_K$ does not mix different valley states. However, the potential term can couple different valley states, when the electrical potential varies abruptly. For example, when an electron is in a donor potential, or an electron is near a rough interface.
We can separate the contribution from donor potential into two terms $V_{jj}$ and $V_{jj^\prime}$, where $V_{jj}$ does not couple and $V_{jj^\prime}$ couples different valley states.

With the knowledge of the orbital Hamiltonian $H_O$, we can evaluate the commutation relation $[r,H_O]$, we will take $[x,H_O]$ as an example. Consider $x\approx \sum_j x^{(j)}\ket{j}\bra{j}$, where $x^{(j)}=\bra{j}x\ket{j}$ is a coordinate operator that does not couple different valley states, and it satisfies commutation relation $[x^{(j)},p^{(j)}] = i\hbar$. Then,
\beq
[x,H_O]=[x, \sum_i \sum_j \frac{(p_i^{(j)})^2}{2m_i^{(j)}}\ket{j}\bra{j}]=i\hbar \sum_j \frac{(p_x^{(j)})^2}{m_x^{(j)}}\ket{j}\bra{j}.
\eeq
Therefore,
\beqa
&&\bra{\bar{0}}[x,H_O]\ket{\bar{r}}=\bra{\bar{0}}xH_O\ket{\bar{r}} - \bra{\bar{0}}H_O x\ket{\bar{r}} \nonumber\\
&=& (E_{\bar{r}}-E_{\bar{0}})\bra{\bar{0}}x\ket{\bar{r}} = i\hbar \sum_j \bra{\bar{0}} \frac{(p_x^{(j)})^2}{m_x^{(j)}}\ket{j}\bra{j}{\bar{r}}\rangle \nonumber\\
&=& i\hbar \sum_{n} C_{\bar{0}n}^* C_{\bar{r}n^\prime} \bra{n}\sum_j \frac{p_x^{(j)}}{m_x^{(j)}}\ket{j}\bra{j}{n^\prime}\rangle \nonumber\\
&=& i\hbar \sum_{nn^\prime} \sum_{j} C_{\bar{0}n}^* C_{\bar{r}n^\prime} \alpha_{n}^{(j)}\alpha_{n^\prime}^{(j)}\bra{F_{nj}} \frac{p_x^{(j)}}{m_x^{(j)}} \ket{F_{n^\prime j}}, \nonumber
\eeqa
while what we need is
\beqa
&&\bra{\bar{0}} p_x \ket{\bar{r}}  = \bra{\bar{0}} \sum_j p_x^{(j)} \ket{j}\bra{j}\ket{\bar{r}} \nonumber\\
&=& i\hbar \sum_{nn^\prime} \sum_{j} C_{\bar{0}n}^* C_{\bar{r}n^\prime} \alpha_{n}^{(j)}\alpha_{n^\prime}^{(j)} \bra{F_{nj}} {p_x^{(j)}} \ket{F_{n^\prime j}}, \nonumber
\eeqa
Thus, there is no direct connection between $(p_x)_{\bar{0}\bar{r}}\equiv \bra{\bar{0}} p_x \ket{\bar{r}}$ and $x_{\bar{0}\bar{r}} \equiv \bra{\bar{0}} x \ket{\bar{r}}$. However, we can estimate a value by using
\beq
(p_x)_{\bar{0}\bar{r}}\approx m^*E_{\bar{r}\bar{0}}x_{\bar{0}\bar{r}}/(i\hbar),
\eeq
where $E_{\bar{r}\bar{0}}=E_{\bar{r}}-E_{\bar{0}}$ is the energy difference of the orbital eigenstates, and the effective mass $m^*$ can be chosen as $m^*=\frac{2m_t + m_l}{3}\approx 0.43 m_0$ or $m^*=3(\frac{2}{m_t} + \frac{1}{m_l})^{-1}\approx 0.26 m_0$, or $m^*=2(\frac{1}{m_t} + \frac{1}{m_l})^{-1}\approx 0.315 m_0$. We can also find out the upper and lower bound values of $\bra{0} p_x \ket{r}$ by using $m^*=m_l=0.92 m_0$ and $m^*=m_t=0.19 m_0$. In our calculation, we choose $m^*= 0.315 m_0$. By using the single effective mass $m^*$, the estimated matrix element can be different from actual values by at most a factor of three.

\section{comparison with bulk donor: ZI}
\label{sec:bulkHz}

In this appendix, we compare the ZI of an electron in a bulk P donor and the ZI in a coupled donor-QD system.

In the case of a bulk P donor in silicon,
we consider the lowest six valley states, i.e. ground state $\ket{A}$ ($n=1$), three-fold degenerate states $\ket{T}$ ($n=2,3,4$), and two-fold degenerate states $\ket{E}$ ($n=5,6$). The electron ZI in a bulk P donor is given by (${r}\ne 0$)
\beq
H_{Z,1{r}} ^{(P)} 
= \frac{1}{2}\mu_B \vec{\sigma}\cdot  \tensor{g}^{(P1,P{r})} \cdot \vec{B},
\eeq
\beq
\tensor{g}^{(P1,P{r})}=g_{ani} \tensor{\Delta}^{(P1,P{r})},
\eeq
\beqa
\tensor{\Delta}^{(Pn,Pn^\prime)}=\sum_{j} a_{P,n}^{(j)}a_{P,n^\prime}^{(j)} \tensor{U}^{(j)}. \label{Deltann}
\eeqa
For an electron in the ground state $\ket{A}$, the ZI only couples to states $\ket{E}$ with opposite spin,
\beqa
\tensor{\Delta}^{(P1,P5)}&=&\sum_{j} a_{P,1}^{(j)}a_{P,5}^{(j)} \tensor{U}^{(j)} \nonumber \\
&=&\frac{1}{3\sqrt{2}} \left[- \tensor{U}^{(x)}- \tensor{U}^{(y)}+ 2 \tensor{U}^{(z)}\right],
\eeqa
\beqa
\tensor{\Delta}^{(P1,P6)}&=&\sum_{j} a_{P,1}^{(j)}a_{P,6}^{(j)} \tensor{U}^{(j)} 
=\frac{1}{\sqrt{6}} \left[\tensor{U}^{(x)} - \tensor{U}^{(y)}\right],
\eeqa
whose matrix form is given explicitly in Table \ref{tableDelta}.

\begin{table}[h!]
\centering
 \begin{tabular}{lll}
 \hline\hline
 bulk-P: &
 $\displaystyle \tensor{\Delta}^{(P1,P5)}=\frac{1}{3\sqrt{2}}\left[\begin{array}{ccc}
 -1& 0 & 0\\
 0& -1 & 0\\
 0& 0 & 2
 \end{array} \right]$
  & \\
  & $\displaystyle \tensor{\Delta}^{(P1,P6)}=\frac{1}{\sqrt{6}}\left[\begin{array}{ccc}
 1& 0 & 0\\
 0& -1 & 0\\
 0& 0 & 0
 \end{array} \right]$
 \\ [0.5ex]
 \hline
 P-QD:&
 $\displaystyle \tensor{\Delta}=\tensor{D}^{(00)}-\tensor{D}^{(11)}=\frac{1}{3}\left[\begin{array}{ccc}
 1& 0 & 0\\
 0& 1 & 0\\
 0& 0 & -2
 \end{array} \right]$
 & \\
 \hline\hline
 \end{tabular}
 \caption{Expressions for tensors $\tensor{\Delta}$ in the Cartesian coordinate system. The constants $a_{n}^{(j)}$ entering $\tensor{\Delta}^{(Pn,Pn^\prime)}$ defined by Eq. (\ref{Deltann}) are given in Appendix \ref{sec:EMA}.} \label{tableDelta}
\end{table}

\begin{table}[h!]
\centering
 \begin{tabular}{llccc}
 \hline\hline
  & $\tensor{\Delta}$ &
 $\tensor{\Delta }_{ZZ}$
 &
 $\tensor{\Delta }_{XZ}$
 &
 $\tensor{\Delta }_{YZ}$\\
 \hline
 bulk-P: & $\sqrt{2}\tensor{\Delta}^{(P1,P5)}$ &
 $-1/3 +\cos^2\theta$ 
  &
  $-\frac{1}{2}\sin2\theta$
  &
  0
 \\ 
 \hline
 bulk-P:& $\sqrt{6}\tensor{\Delta}^{(P1,P6)}$ &
 $\sin^2\theta\cos2\phi$
 &
 $\frac{1}{2}\sin2\theta\cos2\phi$
 &
 $-\sin\theta\sin2\phi$
 \\ 
 \hline
 P-QD:& $\tensor{\Delta}$ &
 $1/3 -\cos^2\theta$ 
  &
  $\frac{1}{2}\sin2\theta$
  &
  0
 \\ 
 \hline\hline
 \end{tabular}
 \caption{Expressions for the components of tensors $\tensor{\Delta}$ in the Cartesian coordinate system, where $Z$-axis is along the applied magnetic field (it is also the spin quantization axis in the lowest order approximation). Here, simplified symbols $\theta=\theta_B$ and $\phi=\phi_B$ are used for the polar and azimuthal angles of the applied magnetic field.} \label{DeltaXYZ} 
\end{table}

In comparison, for a coupled donor-QD system, we have (we choose $\tensor{g}^{(P,j)}=\tensor{g}^{(QD,j)}$ for simplicity)
\beqa
(H_Z)_{\bar{0}\bar{n}} &=& \frac{1}{2}\mu_B \vec{\sigma}\cdot  \tensor{G}^{(\bar{0}\bar{n})} \cdot \vec{B},
\eeqa
\beqa
\tensor{G}^{(\bar{0}\bar{r})} 
&=& g_{ani} C_{\bar{0}0}^*C_{\bar{r}0} \tensor{\Delta}, 
\eeqa
where $\tensor{\Delta}=\tensor{D}^{(00)} - \tensor{D}^{(11)}=\tensor{1}/3 - U^{(z)}$ (matrix form is shown in Table \ref{tableDelta}).

The angular dependence of spin-orbit hybridization due to ZI is determined by the $\tensor{\Delta}_{XZ}$ and $\tensor{\Delta}_{YZ}$ components of tensors $\tensor{\Delta}$ in the $(X,Y,Z)$ coordinate system, where $Z$-axis is along the applied magnetic field (it is also the spin quantization axis in the lowest order approximation). The corresponding expressions can be  obtained as shown in Table \ref{DeltaXYZ}.

\section{comparison with bulk donor: Electron-phonon interaction}
\label{sec:bulkHep}

In this appendix, we compare the electron-phonon interaction of an electron in a bulk P donor and electron-phonon interaction in a coupled donor-QD system.

In the case of a bulk donor in silicon, we consider again the lowest six valley states, i.e. $\ket{A}$ ($n=1$), $\ket{T}$ ($n=2,3,4$), $\ket{E}$ ($n=5,6$). Suppose there is an electron in the donor ground state $\ket{A}$, then, the electron-phonon interaction $H_{EP}$ could couple the ground orbital state $\ket{A}$ to states $\ket{E}$, and the electron-phonon interaction of an electron in a bulk donor in silicon is given by (${r}\ne 0$)
\beq
(H_{EP}^{(P)})_{{1}{r}}^{\uparrow\uparrow} = \sum_{\vec{q}\lambda} M_{\vec{q}\lambda,{1}{r}}^{(P)} (b_{-\vec{q}\lambda}^\dag + b_{\vec{q}\lambda}),
\eeq
\beq
M_{\vec{q}\lambda,1{r}}^{(P)} =\sum_{j} a_{1}^{(j)}a_{{r}}^{(j)}f_{1{r} }^{(j)}(\vec{q}) M_{\vec{q}\lambda}^{(j)},
\eeq
where in the limit of long wave phonons, we have $f_{1r}^{(j)}(\vec{q})\approx \bra{F_{1}^{(j)}}1\ket{F_{1}^{(j)}}\approx 1$.
Therefore, 
\beqa
M_{\vec{q}\lambda,1{r}}^{(P)} &=& \sum_j a_{1}^{(j)} a_{{r}}^{(j)} M_{\vec{q}\lambda}^{(j)} 
=i\sqrt{\frac{\hbar q}{2\rho_c v_{\lambda}}}\Pi_{\vec{q}\lambda,1{r}}^{(P)},
\eeqa
\beqa
{\Pi}_{\vec{q}\lambda,1{r}}^{(P)}
&=&\Xi_{u}  \hat{e}^{(\vec{q}\lambda)} \cdot \tensor{\Delta}^{(P1,P{r})}\cdot \hat{q},
\eeqa
where the expressions for the tensors $\tensor{\Delta}^{(P1,P{r})}$ are listed in Table \ref{tableDelta}.

In comparison, for an electron in the coupled donor-QD system, the electron-phonon interaction is
\beq
(H_{EP})_{\bar{0}\bar{r}}^{\uparrow\uparrow} = \sum_{\vec{q}\lambda} (M_{\vec{q}\lambda})_{\bar{0}\bar{r}} (b_{-\vec{q}\lambda}^\dag + b_{\vec{q}\lambda}),
\eeq
\beq
(M_{\vec{q}\lambda})_{\bar{0}\bar{r}}
= iC_{\bar{0}0}^*C_{\bar{r}0}\sqrt{\hbar q/2\rho_c v_{\lambda}}{\Pi}_{\vec{q}\lambda}^\prime ,
\eeq
\beqa
{\Pi}_{\vec{q}\lambda}^\prime =\Pi_{\vec{q}\lambda,00} - \Pi_{\vec{q}\lambda,11}
&=&\Xi_{u}\hat{e}^{(\vec{q}\lambda)} \cdot \tensor{\Delta} \cdot \hat{q},
\eeqa
where $\Pi_{\vec{q}l}^{\prime} = \Xi_u (1/3-\cos^2\vartheta)$, $\Pi_{\vec{q}t_1}^{\prime}= \Xi_u \cos\vartheta\sin\vartheta$, and $\Pi_{\vec{q}t_2}^{\prime}=0$.

The difference of electron-phonon interaction will modify the angular distribution of phonon emissions. Thus, it will modify the magnitude of spin relaxation after averaging phonon modes in all three dimensions.

\section{$\theta_B$, $\phi_B$ dependencies}

\begin{figure}
\includegraphics[scale=0.9]{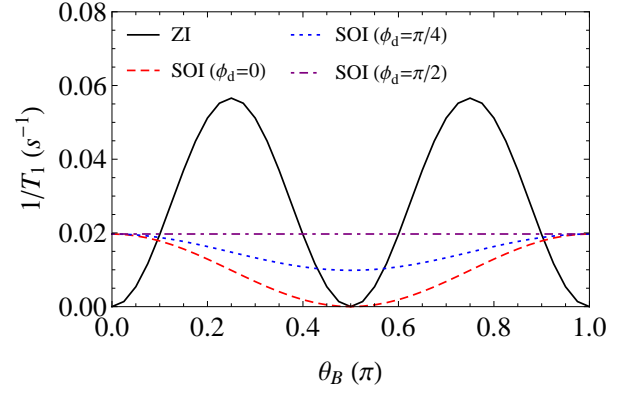}
\caption{Spin relaxation as a function of the polar angle $\theta_B$ of the applied magnetic field when detuining $\epsilon=0$ meV and $B=0.5$ T. We show spin relaxation due to ZI (black solid line) with $\theta_B=\pi/4$ and $\phi_B=0$, and spin relaxation due to SOI when $d_{\parallel}=2$ nm with $\phi_B-\phi_d=0$ (red dashed line), $\phi_B-\phi_d=\pi/4$ (blue dotted line), and $\phi_B-\phi_d=\pi/2$ (purple dot-dashed line).}\label{Fig_3}
\end{figure}

\begin{figure}
\includegraphics[scale=0.9]{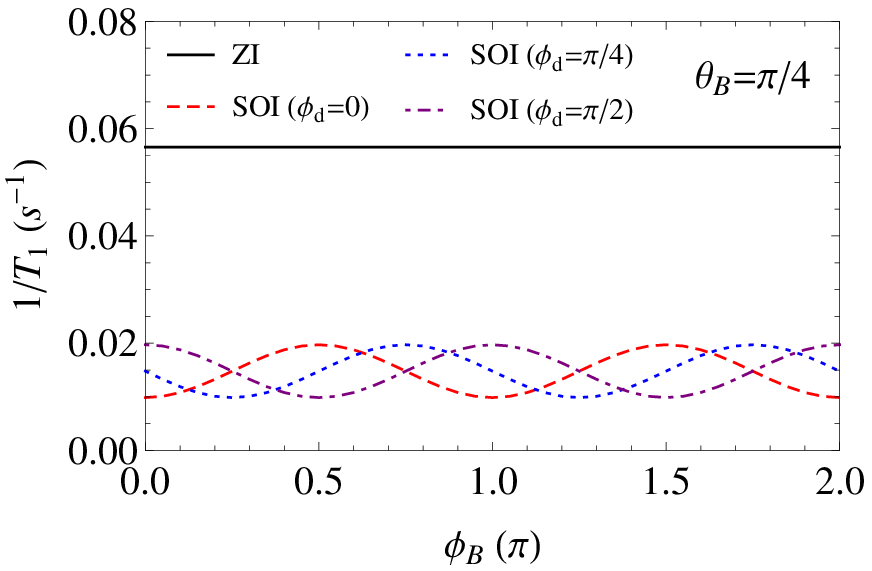}
\caption{Spin relaxation as a function of the azimuthal angle $\phi_B$ of the applied magnetic field when detuining $\epsilon=0$ meV, $B=0.5$ T and $\theta_B=\pi/4$. We show spin relaxation due to ZI (black solid line), and spin relaxation due to SOI when $d_{\parallel}=2$ nm with $\phi_d=0$ (red dashed line), $\phi_d=\pi/4$ (blue dotted line), and $\phi_d=\pi/2$ (purple dot-dashed line).}\label{Fig_4}
\end{figure}

\begin{figure}
\includegraphics[scale=0.9]{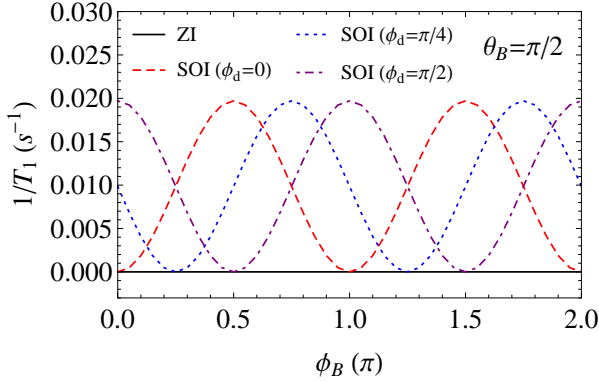}
\caption{Spin relaxation as a function of the azimuthal angle $\phi_B$ of the applied magnetic field when detuining $\epsilon=0$ meV, $B=0.5$ T and $\theta_B=\pi/2$. We show spin relaxation due to ZI (black solid line), and spin relaxation due to SOI when $d_{\parallel}=2$ nm with $\phi_d=0$ (red dashed line), $\phi_d=\pi/4$ (blue dotted line), and $\phi_d=\pi/2$ (purple dot-dashed line).}\label{Fig_5}
\end{figure}

Besides the dependence on the magnitude of the applied magnetic field, we can also study the dependence on the orientation of the applied magnetic field. 
We only report the angular dependence for the inter-donor-QD spin relaxation here.

Figure \ref{Fig_3} shows the spin relaxation rate $1/T_1$ due to each mechanism as a function of the polar angle $\theta_B$ of the applied magnetic field when detuning $\epsilon= 0$ meV and $B=0.5$ T. We show spin relaxation due to ZI when $g_{ani}=0.001$ (black solid line), and spin relaxation due to SOI when $d_{\parallel}=2$ nm (red dashed line). We choose the azimuthal angle $\phi_B=0$ for spin relaxation due to ZI (black solid line), and $\phi_B-\phi_d=0$, $\phi_B-\phi_d=\pi/4$ and $\phi_B-\phi_d=\pi/2$ for spin relaxation due to SOI. The spin relaxation due to ZI goes to zero when the polar angle $\theta_B=0$ or $\pi/2$, i.e. when the magnetic field is in-plane or out-of-plane, and it is maximum when $\theta_B=\pi/4$. The spin relaxation due to ZI vanishes at certain orientations of magnetic field due to the vanishing of hybridization as discussed above. However, spin relaxation due to SOI becomes maximum when $\theta_B=0$ (in-plane) and minimum when $\theta_B=\pi/2$ (out-of-plane).

Figure \ref{Fig_4} shows the spin relaxation rate $1/T_1$ due to each mechanism as a function of the azimuthal angle $\phi_B$ when detuning $\epsilon= 0$ meV, $B=0.5$ T and $\theta_B=\pi/4$.
We use the same $g_{ani}$ for ZI mechanism and the same $d_{\parallel}$ for the SOI mechanism as in Figure \ref{Fig_3}.
We choose $\phi_d=0$, $\phi_d=\pi/4$ and $\phi_d=\pi/2$ for spin relaxation due to SOI. Since the spin relaxation due to ZI is maximized when $\theta_B=\pi/4$, it dominates over the spin relaxation due to SOI. As shown in the figure, the spin relaxation due to ZI shows no dependence with $\phi_B$.

Figure \ref{Fig_5} shows the spin relaxation rate $1/T_1$ due to each mechanism as a function of the azimuthal angle $\phi_B$ when detuning $\epsilon= 0$ meV, $B=0.5$ T and $\theta_B=\pi/2$.
We use the same $g_{ani}$ and $d_{\parallel}$ as in Figure \ref{Fig_3} and \ref{Fig_4}.
The spin relaxation due to ZI is suppressed when $\theta_B=\pi/2$, and the spin relaxation due to SOI dominates. We choose $\phi_d=0$, $\phi_d=\pi/4$ and $\phi_d=\pi/2$ for spin relaxation due to SOI. Spin relaxation due to SOI depends on $\phi_B-\phi_d$, as previously indicated.
The rate is minimum when $\phi_B=\phi_d$ and maximum when $\vec{B}$ and $\vec{d}$ are orthogonal.
For in-plane B, the relaxation can be completely suppressed when $\phi_B=\phi_d$.
By changing the angle of the QD shift relative to the donor with an electric field, we can modify the  dependence of spin relaxation with azimuthal angle $\phi_B$ of the applied magnetic field. Thus, the electric field (in-plane) dependence of spin relaxation can be applied to tell whether the relaxation is dominated by ZI or SOI.


\bibliographystyle{apsrev4-1-nourl}
\bibliography{donordot}

\end{document}